\documentclass[final,5p,times,twocolumn]{elsarticle}

\usepackage{graphicx}
\usepackage{amssymb}
\usepackage{lineno}
\usepackage{bm}
\usepackage{amsmath}

\begin{document}

\begin{frontmatter}

\title{Machine Learning of Two-Dimensional Spectroscopic Data}

\author[1]{Mirta Rodr\'iguez}
\ead{rodiguez@zib.de}
\author[1,2]{Tobias Kramer}
\ead{kramer@zib.de}
\address[1]{Zuse Institute Berlin (ZIB), Takustr.\ 7, 14195 Berlin, Germany}
\address[2]{Department of Physics, Harvard University, 17 Oxford Street, 02138 Cambridge, Massachusetts, United States}

\begin{abstract}

Two-dimensional electronic spectroscopy has become one of the main experimental tools for analyzing the dynamics of excitonic energy transfer in large molecular complexes. Simplified theoretical models are usually employed to extract model parameters from the experimental spectral data.  Here we show that computationally expensive but exact theoretical methods encoded into a neural network can be used to extract model parameters and infer structural information such as dipole orientation from two dimensional electronic spectra (2DES) or reversely, to produce 2DES from model parameters. 
We propose to use machine learning as a tool to predict unknown parameters in the models underlying recorded spectra and as a way to encode computationally expensive numerical methods into efficient prediction tools. We showcase the use of a trained neural network to efficiently compute disordered averaged spectra and demonstrate that disorder averaging has non-trivial effects for polarization controlled 2DES. 
\end{abstract}


\end{frontmatter}
\section{Introduction}
Spectral data resulting from the interaction of light with matter is used as a diagnostic tool in many scientific and technological areas including physics, astrophysics, chemistry, or biology. Common to all spectroscopic techniques is the amount of data generated and the requirement of several fitting models and parameters for understanding the underlying time-dependent mechanisms of the observed physical systems. Many of those parameters, such as the dipole moment in molecular spectra are difficult to obtain from first principles. This makes the use of data-based algorithms a traditional technique in molecular spectroscopy \cite{Adolphs2006a,Micaelli2017,Roeding2017a}.

Light harvesting complexes (LHCs) are  pigment-protein molecular systems, which are part of the photosynthetic apparatus of bacteria and green plants \cite{Blankenship2014,Chmeliov2016}.
Absorbed photons create an electronic excitation in the pigment, the exciton,  which is transmitted through the light harvesting complex due to dipole-dipole interactions. Proteins surrounding the pigments are treated as an external environment for the exciton that results in energy dissipation into pigment vibrations. Energy transfer mechanisms in LHCs have been understood during the last decades with the use of linear spectra and more recently with time-dependent techniques such as pump-probe and two-dimensional electronic spectroscopy (2DES). 

In order to reproduce experimental data, computationally expensive methods such as the exact hierarchical equations of motion (HEOM) formalism \cite{Tanimura1989,Ishizaki2009,Kreisbeck2011} that considers the exciton and the vibrational environment on equal footing need to be used. Moreover, to reproduce experimental spectra, repeated computations are required for accounting for the different instantaneous exciton energies due to the slow protein movement, which is called static disorder. As an example, the computation of the 2DES of the paradigmatic Fenna Matthews Olson complex (FMO) for an arbitrary polarisation laser pulse sequence requires $21$ calculations of $6$ pathways to account for rotational averaging, each calculation consisting of several hundred time-propagations. All these computations have to be repeated over hundreds of disorder realizations \cite{Hein2012}. In practice simplified models, such as the Redfield approximation valid for weak environmental couplings, are commonly used in spectra fits. The effect of disorder is usually simplified to Gaussian convolutions of specific widths \cite{Somsen1996}.
\begin{figure}[htbp]
\centering
\includegraphics[draft=false,width=0.45\textwidth]{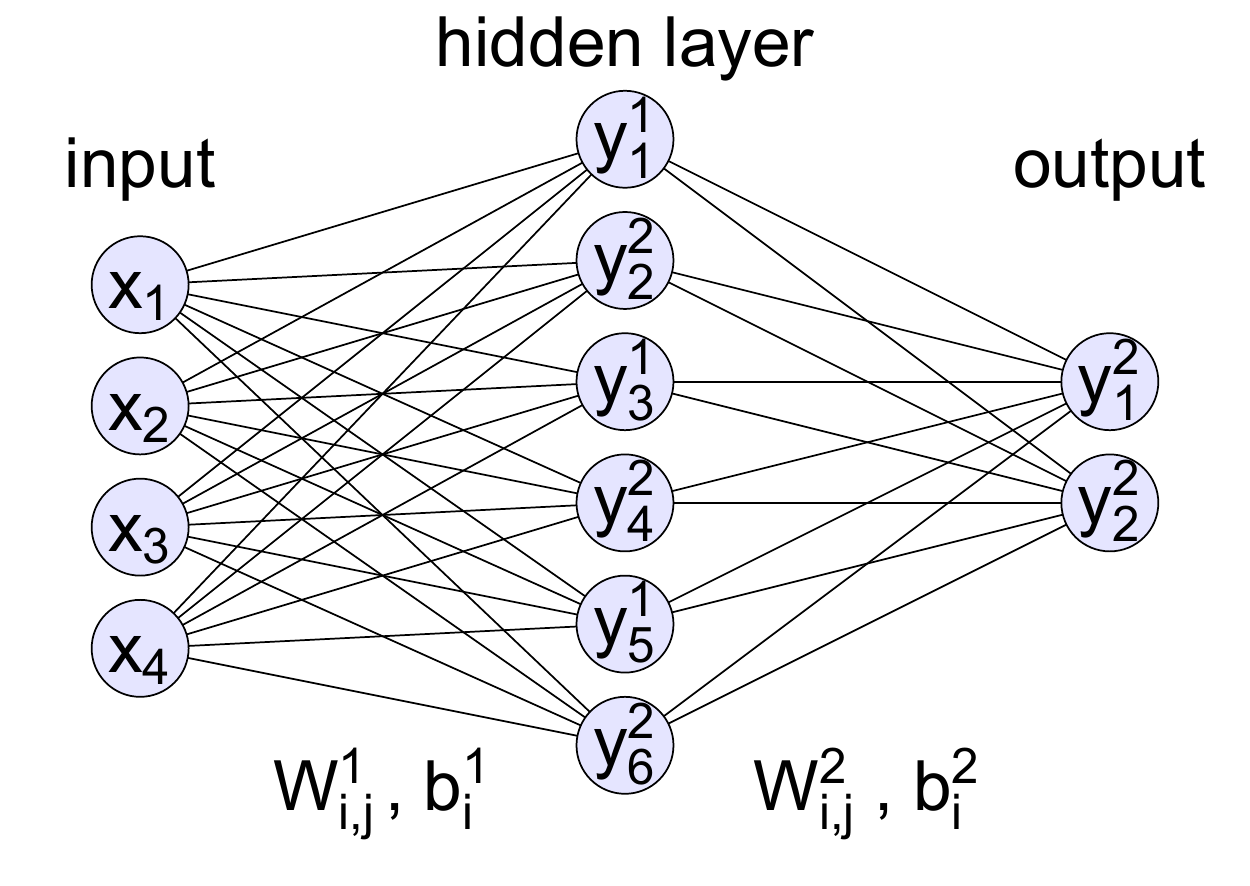}
\caption{Scheme of 2-layer neural network with one hidden layer. Fitting parameters are encoded in the transformation matrices $\mathbf{W}$, $\mathbf{b}$}
\label{fig:2NN}
\end{figure}
Machine learning (ML) is a long standing technique \cite{Lecun1998,Hinton2006,Krizhevsky2012,Goodfellow2016} for the extraction of patterns from large amounts of data that is now being explored as a tool to solve outstanding problems in quantum physics. First results have been obtained using neural network (NN) algorithms for translational invariant many-body systems where finite correlation lengths underlay in the data \cite{VanNieuwenburg2016, Carrasquilla2016,Mavadia2016,Carleo2017}. Time dependent transport problems have been also studied using ML \cite{Hase2017,Bandyopadhyay2018,Dral2018}. Time propagation in LHC requires more involved neural network configurations and data sampling techniques \cite{Hase2017}. Moreover, ML is being discussed as a technique to bypass expensive calculations in quantum chemistry with pre-trained networks \cite{Rupp2012,Montavon2013,Duvenaud2015,Rupp2015a}.

In this paper we show that ML is an efficient tool to predict unknown parameters in the models underlying recorded spectra and reversely, to produce accurate 2DES spectra from given parameters. The machine learning model is based on 
artificial neural network algorithms (see fig.~ \ref{fig:2NN}), that exploit the underlying correlations between the
spectra and the molecular properties. We demonstrate two applications and discuss the optimization of the underlying network in terms of network layout and number of parameters. First, we present a machine learning model, trained on a database of exact calculation results for
thousands of 2D electronic spectra, that predicts with high accuracy the orientations of the dipole moments in FMO.
We attain a standard deviation of only $0.01^\circ$ over a testset not seen by the network.
This uncertainty is two orders of magnitude better than current theoretical and experimental fits. 
Next we apply a neural network algorithm to predict 2DES from model parameters and study the effect of disorder in 2DES.
Inspired by recent experimental results \cite{Thyrhaug2018} we have calculated 2DE spectra for a non-trivial laser  polarisation sequence $\langle 45^\circ,-45^\circ,90^\circ,0^\circ \rangle$.
In order to reproduce the experimental spectra we need to perform disorder averaging of our calculations. We show that a trained NN can efficiently reproduce computationally expensive 2DES calculations from only seven parameters. This is an efficient method to store and distribute computationally expensive calculations. 
 
All neural networks are trained from a data set of 2D spectra images calculated with the distributed hierarchical equations of motion (DM-HEOM) \cite{Noack2018,Kramer2018a},
which provides an efficient and scalable implementation of HEOM.
The neural networks, the training data, and the trained networks are available in the Supplementary Material.

\section{Data generation: 2DES}

One of the most studied light harvesting complexes is the FMO complex \cite{Fenna1975,Olson2004} which acts as an excitonic wire in the photosynthetic apparatus of green sulfur bacteria, channeling the energy from the antenna to the reaction center. 
Since it was the first photosynthetic complex of which the X-ray structure became available, it has been subject to a wide variety of theoretical and spectroscopic studies and became one of the best characterized pigment–protein complex, see e.g. \cite{Milder2010}.

The energy transport in the LHC photosynthetic systems is modeled using the Frenkel exciton description \cite{May2004}, where the energy channels across a discrete network with on-site energies and off-diagonal couplings between the pigments,  coupled to a vibrational bath. The external electromagnetic field from the probing laser pulses 
is usually treated in the impulsive limit \cite{Mukamel1995}. Site energies and pigment-pigment couplings $\varepsilon_{ab}$ cannot be deduced from optical experiments directly and are usually obtained by fits to optical spectra together with first principles calculations \cite{Adolphs2006a}. Exciton simulations, including effective pigment-protein interactions are used to generate spectra and to compare to experimental results \cite{Adolphs2006a}. Important parameters in these simulations are the dipolar coupling strength and the line width of the transitions, related to the pigment-protein interactions.

Two-dimensional electronic spectroscopy (2DES) is an experimental technique with high temporal and spectral resolution that has been successfully applied to study the energy-transfer pathways in photosynthetic complexes, in particular for the FMO system \cite{Brixner2005,Engel2007a}. In contrast to absorption or fluorescence measurements, 2DES  provides the full correlation map between the excitations and the probing wavelengths as a function of time after initial light absorption \cite{Mukamel1995,Cho2005,Nuernberger2015}. 
In 2DES a sequence of three laser pulses creates coherences between the ground state and between the exciton states, which are read out by a fourth pulse. The first and last time intervals between the pulses are converted by a Fourier transform to the frequency domain, with $\omega_1$ and $\omega_3$ (see Fig.~\ref{fig:TD}) denoting the excitation and
 emission frequencies, respectively. The remaining interval between second and third pulse sets the delay time $\tau$. As any other molecular spectra, the  2DES peaks are masked by the dipole moments of the pigments and their relative orientations.
The dipole moments of the pigments reflect the capacity of each pigment to absorb light due to the electron charge distribution and the relative orientation to the incoming light. This is encoded in the dipole moment $\hat{\mu}$ operator
$\hat{\mathbf \mu}= \hat{\mathbf \mu}^++\hat{\mathbf \mu}^-$,  where
\begin{equation}
\hat\mu^+=\sum_{a=1}^{N_{\rm pigments}}  \mathbf{d}_a |a\rangle\langle 0|\,,
\label{eq:mu_plus}
\end{equation}
and $\mathbf{d}_a=q \mathbf{r}$ is the total charge times the direction from positive to negative charge.
Lacking conclusive results from challenging first principle calculations it is a general consensus in the field to assign the dipole orientations of the BChls to the $N_B-N_D$ axis, although deviations of up to $6^\circ$ have been pointed out by several authors \cite{Cole2013,Muh2007,Adolphs2008a}.

We use the distributed DM-HEOM \cite{Noack2018,Kramer2018a} to efficiently generate the 2DES data.
A complete expression of the 2D spectra in terms of the dipole moment operators can be found in \cite{Kramer2018a}, Eqs.~(64-69).
We compute the rephasing pathways (RP) and non-rephasing (NR), ground state bleach (GB), stimulated emission (SE) and excited state absorption (ESA) at $T=100$~K and several delay times $\tau$. Besides the seven single exciton states of the FMO system \cite{Adolphs2006a}
\begin{equation}
 \varepsilon=
  \left({
  \begin{array}{ccccccc}
   12410 & -87.7 & 5.5  & -5.9 & 6.7 & -13.7 & -9.9 \\
   -87.7 & 12530 & 30.8 & 8.2 & 0.7 & 11.8 & 4.3 \\
    5.5 & 30.8 & 12210 & -53.5 & -2.2 & -9.6 & 6.0 \\
    -5.9 & 8.2 & -53.5 & 12320 & -70.7 & -17.0 & -63.3 \\
     6.7 & 0.7 & -2.2 & -70.7 & 12480 & 81.1 & -1.3 \\ 
     -13.7 & 11.8 & -9.6 & -17.0 & 81.1 & 12630 & 39.7 \\ 
     -9.9 & 4.3 & 6.0 & -63.3 & -1.3 & 39.7 & 12440      
   \end{array} }
   \right)\,\text{cm$^{-1}$.}
\label{eq:Hamiltonian-site-basis-H-lambda}
\end{equation}
, additionally $28$ double exciton states are explicitly included in the Hamiltonian and in the dipole matrix \cite{Cho2005,Hein2012}.
For the vibrational couplings of the pigments, we consider seven independent sets of harmonic oscillators coupled linearly to each pigment. The spectral density of the vibrations of each pigment is given by Drude-Lorentz shape $J(\omega)=2\lambda\frac{\omega\gamma}{\omega^2+\gamma^2}$ with a correlation time $\gamma^{-1}=50$~fs and $\lambda=35$~cm$^{-1}$. 

\subsection{Dipole orientation}
\begin{figure}[htbp]
\centering
\includegraphics[draft=false,width=0.45\textwidth]{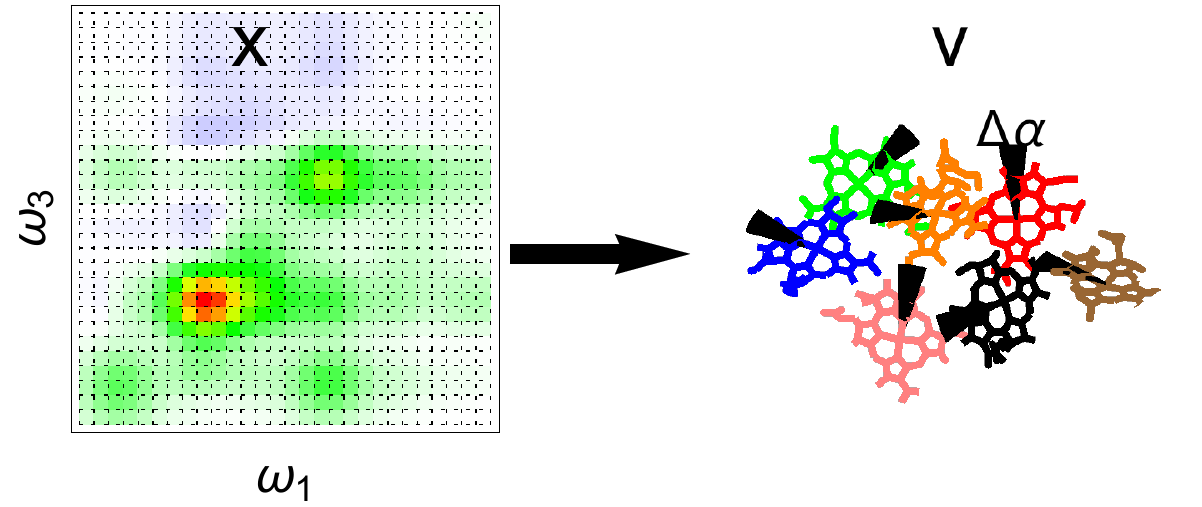}
\caption{We perform supervised learning experiments with $10,000$ data sets $\{\mathbf{x}\rightarrow\mathbf{v}\}$, where each image $\mathbf{x}$ is assigned to a $7$ dimensional real valued vector $\mathbf{v}=\Delta \alpha$ that denotes the angular deviation of the FMO pigment orientations from the $N_D-N_B$ axis. Input images $\mathbf{x}$ are $28\times28$ pixel 2DES of FMO at $\tau=200$ fs calculated using HEOM for a linearly polarised light sequence. Calculations include all GB, SE and ESA RP and NR pathways. }
\label{fig:TD}
\end{figure}

We produce $10,000$ 2DE spectra for a sequence of linearly polarized light for $7\times 10,000$ different random pigment orientations $\mathbf{d}_a$ at a fixed delay time $\tau=200$ fs.
To constrain the parameter space, we allow only for rotations of up to $\pm 10^\circ$ of the dipole vectors within the plane defined by the $N_B,N_C,N_D$ atoms in each bacteriochlorophyll, see Fig.~\ref{fig:TD}.
The angle $\Delta \alpha \subset$ ($-10^\circ$,$+10^\circ$) denotes the angular deviation from the $N_B,N_D$ direction in this plane.

\subsection{Sequences of polarized lasers}
We produce 2DE spectra for a laser field polarisation sequence $\langle 45^\circ,-45^\circ,90^\circ,0^\circ\rangle$ using Eqs.~(71)-(74) in \cite{Kramer2018a}.
In order to reproduce experimental results from \cite{Thyrhaug2018} we need to add static disorder $\Delta \mathrm{\varepsilon}$ to the diagonal of the seven site Hamiltonian $\varepsilon_{ab}$.
At four delay times $\tau$, we draw $5,000$ random realisations of static disorder $\Delta \varepsilon$ from a Gaussian distribution of standard deviation (STD) $50$~cm$^{-1}$ for the diagonal of the FMO Hamiltonian $\varepsilon_{aa}$ and compute the corresponding 2DES (see Fig.~\ref{fig:TDIS}). For comparison with experiments, we discuss only the measured rephasing signals.

\section{Machine learning method}
\label{sebsec:mlearning}

\subsection{Neural Network algorithm}

One of the most common techniques in ML is the use of a neural network, which is an algorithm based on a collection of interconnected units called artificial neurons that are inspired by axons in a biological brain. Neurons are organised in layers which perform linear and non-linear transformations to the input signal that is propagated to the output layer (see Fig.~\ref{fig:2NN}). Common non linear operations include projection, pooling and convolution.

For a $M$-layer network, the network response $\bf{y}^m$ at layer $m=1,\dots,M$ is computed as
\begin{eqnarray}
    h_{i}^m &=&\sum_{j=1}^{N_h^{m-1}} W^{m}_{i,j}y_j^{m-1}+b_i^m, \\
    y_i^m &= & \sigma^m(h_i^m),
    \label{eq:NN}
\end{eqnarray}
where $i=1,..,N_h^m$ and $N_h^m$ is the number of neurons in the $m$-th hidden layer and the input vector $\bf{y}^0=\bf{x}$ of dimension $N_h^0=N_{\rm in}$. The network output of dimension $N_{\rm out}$ is given by $\mathbf{y}^M=\mathbf{W}^{M}\mathbf{y}^{M-1}+\mathbf{b}^M$. The linear operation at each layer $m$ is encoded in the weight ($\mathbf{W}^m$) and bias ($\mathbf{b}^m$) matrix. The activation function $\sigma^m$ in the hidden layers is commonly the $\tanh$ function or ramp$(x)$ that gives $x$ if $x \geq 0$ and $0$ otherwise, or the linear function for the output layer $m=M$.  

In the supervised learning approach training the NN algorithm means finding the optimal parameters $\mathbf{W}^m,\mathbf{b}^m$ that minimize the loss function 
\begin{equation}
     \mathcal{L} = \frac{1}{N} \sum_{l=1}^N (\mathbf{v}_l - \mathbf{y}_l(\mathbf{x}_l))^2,
     \label{eq:loss}
\end{equation}
that represents the error of the output $\mathbf{y}$ calculated over a training data set $\{\mathbf{x}\rightarrow \mathbf{v}\}_l$ of $N$ samples, where each input $\mathbf{x}$ is assigned to a label $\mathbf{v}$ which is the actual value used in the calculation or could be the settings used in a controlled experiment. The network is trained to approximate the output function $\bf{y}=\bf{y}^M$,  Eq.~(\ref{eq:NN}) provided by the NN algorithm and its partial derivatives with respect to each of the elements of the input vector $\mathbf{x}$. If the data set $N$ is large the evaluation of the derivatives to minimize the sum Eq.~(\ref{eq:loss}) at every step becomes too costly and the Stochastic Gradient Descend (SGD) method is used. With SGD the gradient estimation over random data batches of fixed size is computed and the minimum is found by iterating the gradient estimation over the data set. A full iteration over the dataset is called an epoch. Advanced SGD methods speed up convergence towards a minimum by adding momentum. All the results presented here are calculated using Wolfram Mathematica 11.3 Machine Learning capabilities which are based on the MXNet Learning Framework using NVIDIA GPUs. 

We randomly divide the available data into training, validation, and test data, and calculate the performance of the trained neural network on test data which was not used during the training.
We find similar training results when keeping the ratio
\begin{equation}
\frac{\text{batch size~}B}{\text{size of training set~}N}
\end{equation}
constant.
In all the results shown here we have fixed the batch size $B=100$ and the number of epochs to $100,000$ and use $N=9,000$ (dipole) $N=4,500$ (disorder) training data. We use renormalised data both for the input $\mathbf{x}$ and label data $\mathbf{v}$ when training the NN.  In all our computations we used the ADAM SGD minimisation method \cite{Kingma2014} with fixed learning rate $0.001$ and $\beta_1=0.9$ and $\beta_2=0.999$. 

A seminal paper \cite{Hornik1989} shows that multilayer feedforward networks such as those in Eqs.~(\ref{eq:NN}) with as few as one hidden layer are capable of universal approximation.
Over the years more complex networks emerged. One example is the
Long Short Term Memory (LSTM) recurrent neural network that contains a non-linear recurrent memory cell which allows for retaining information across the data in a non trivial way \cite{Hochreiter1997}. As we show here, LSTM type networks routinely outperform simpler networks without introducing more parameters. Given that we perform image based training we also test the use of convolutional neural network algorithms \cite{Lecun1998} which have had a great success in image recognition, with alternating pooling (subsampling) and convolution layers. The detailed layout, number of layers and parameters, and training evolution is shown in the \ref{sec:AppendixNN}.

\subsection{Prediction of the orientations of the FMO complex dipole moments}

The calculation of the strength and orientation of the dipole moment of the constituent pigments of LHCs is a very difficult task due to screening effects. A commonly adopted approach is to align the  $Q_y$ transition dipole to the $N_B-N_D$ axis of the isolated BChl a pigments, although deviations are expected due to the surrounding proteins. 
Deviations of up to $7^\circ$ result in better fit of Circular Dichroism spectra \cite{Adolphs2008a} and first principles calculations based on transition charge from electrostatic potential  \cite{Muh2007} predict orientations within $\pm 2^\circ$ of the $N_B-N_D$ axis.
Also recent ab-initio calculations show a deviation of the dipole orientation of $4^\circ-6^\circ$ with respect to the $N_B-N_D$ axis (\cite{Cole2013} table S3).

We train a NN algorithm on 2DES calculated for dipole orientations of up to $\pm 10^\circ$ with respect to the $N_B-N_D$ axis and use the trained NN algorithm to predict the orientation of 2DES not seen by the algorithm. We produce $10,000$ data sets consisting of $\mathbf{x}_l:$ 2DES $28\times28$ pixel images assigned to a $\mathbf{v}_l:$ dipole orientation $ \mathrm{\Delta \alpha} \in \mathcal{R}^7$, a $7$ dimensional vector of real numbers $\subset$ ($-10.0$, $+10.0$). We randomly divide the available data into training ($9,000$), validation ($500$) and test data ($500$) sets. We feed in the training data into different NN algorithms (see Appendix \ref{sec:DIPOLE}).

For the 2-layer NN algorithm (Fig.~\ref{fig:2NN},\ref{sec:2NN}), the 2DES matrices are reshaped into $N_{\rm in}= 28\times 28$ dimensional input vectors $\mathbf{y}^0=\bf{x}$, which are transformed by the hidden layer of dimension $N_h=50$ into
\begin{eqnarray}
y_i^1 &= & \textrm{ramp} \left(\sum_{j=1}^{N_{\rm in}} W^{1}_{i,j} y_j^{0}+b_i^1\right). \nonumber \\ 
    i & = & 1,\dots,N_h
    \label{eq:2N}
\end{eqnarray}
The network output of dimension $N_{\rm out}=7$ is given by $\mathbf{y}^2=\mathbf{W}^{2}\mathbf{y}^{1}+\mathbf{b}^2$. We renormalise each of the images $\mathbf{x}_l$ by a common value (maximum value of all the images) and the orientation vectors $\mathbf{v}_l: \Delta \alpha$ are divided by $10$. We calculate the loss function Eq.~(\ref{eq:loss}) over the training data (see orange line in Tab.~\ref{tab:2NNdipole}a) and the validation data set (blue line) in order to monitor overfitting. We fix the number of times (epochs$=100,000$) that the algorithm as seen the training data in batches of $100$. This takes $2.3$~h on a NVIDIA GeForce GTX 1080 Ti with 11 GB GDDR5X-RAM.

\begin{figure}[htbp]
 \centering
 \begin{tabular}{cc}
(a)\includegraphics[draft=false,width=0.2\textwidth]{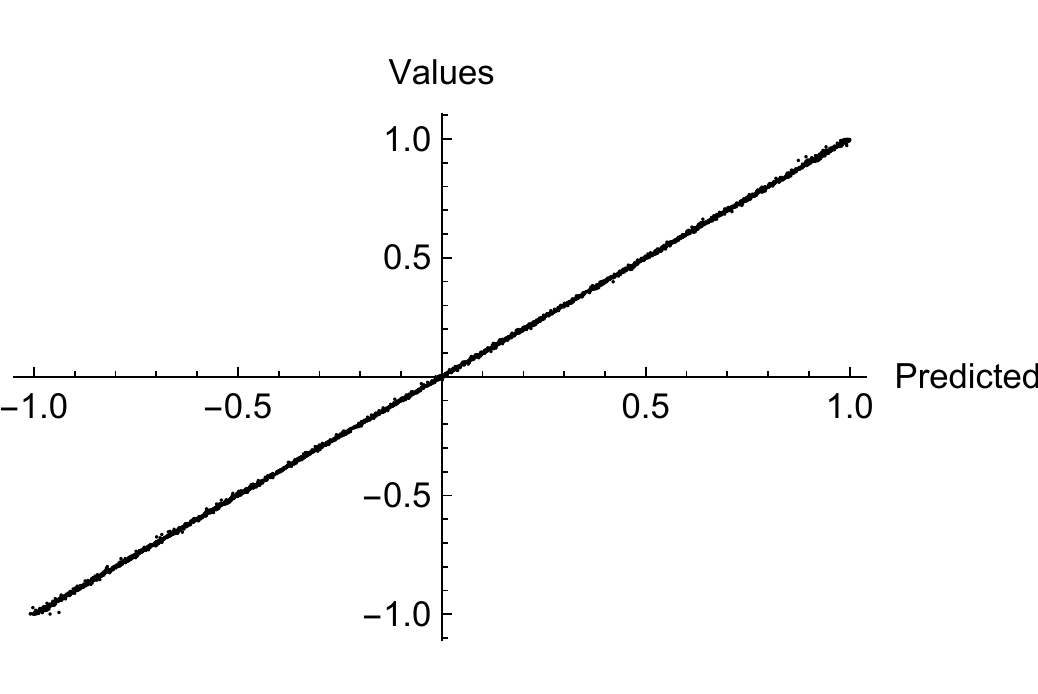} &
(b)\includegraphics[draft=false,width=0.2\textwidth]{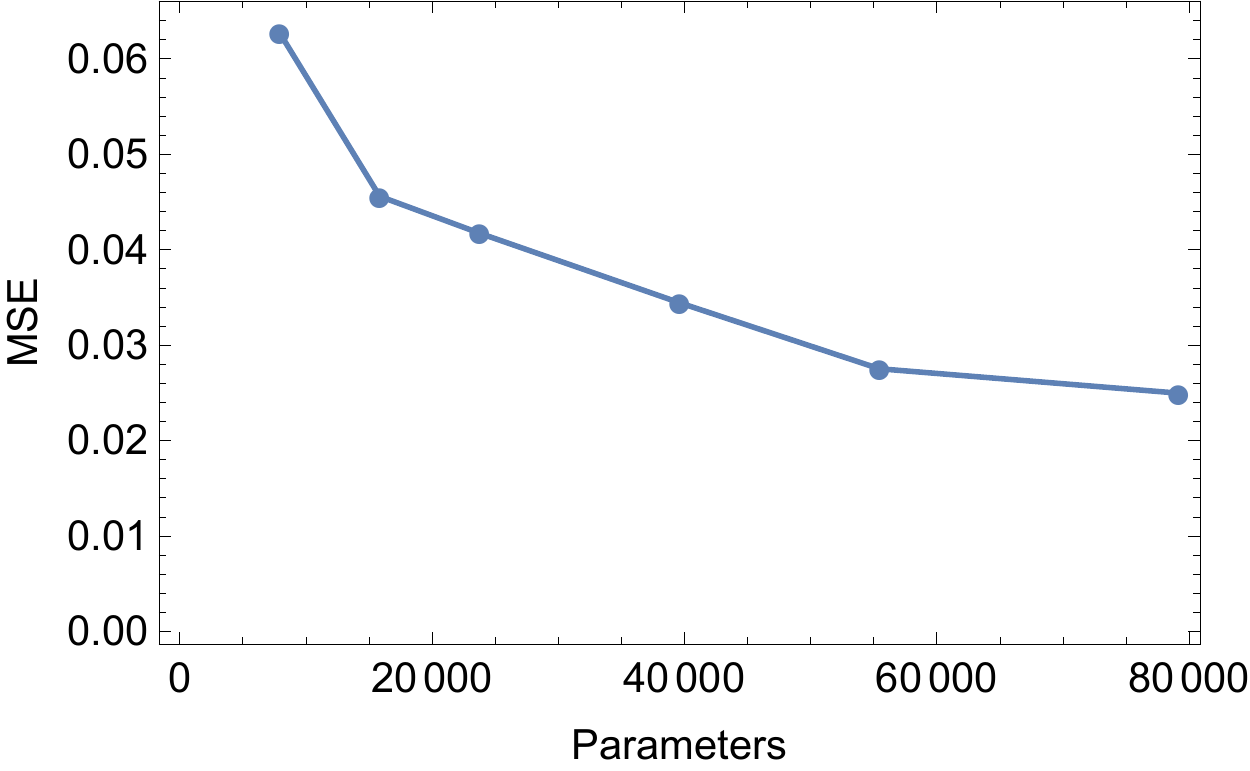}
    \end{tabular}
\caption{(a) Predicted $\mathbf{y^2}$ versus exact normalised values $\mathbf{v}=\Delta \alpha/10$ for a 2-layer NN trained over the FMO pigment dipole orientation from input 2DES (see  \ref{sec:2NN}) for a test set of $500$ random values of the FMO pigment orientation not included during the training of the algorithm.(b)Mean Squared Error (MSE) as a function of fitting parameters for a 2-layer NN algorithm in Eq.~(\ref{eq:2N}) with increasing size of the hidden layer $N_h=\{ 10,20,30,50,70,100\}$. }
\label{fig:predict}
\end{figure}

 A comparison of the predicted dipole orientation values using the algorithm $\mathbf{y}$ versus the actual values $\mathbf{v}$ used in the spectra calculation is showed in Fig.~\ref{fig:predict} a). We have used the small test data set of $500$ spectra which was not used during the training. To calculate the performance of the trained neural network we calculate the standard deviation of the predicted minus actual values of the dipole orientations. We denote by the mean squared error (MSE) the average of this $7$-dimensional vector. We show the MSE for increasing number of neurons in the hidden layer as a function of free parameters in Fig.~\ref{fig:predict} b). 

Results for this simple 2-layer NN and other NN algorithms discussed in \ref{sec:AppendixNN} are summarised in Table~\ref{tab:DIP}. We show in the last column the results for the predicted dipole moment orientation of the FMO reference case, with all dipoles aligned to the $N_B-N_D$ axis. Values obtained are within the MSE of the NNs.

\begin{table*}[t]
{\small
\centering
\begin{tabular}{l|r|r|l}
Network & Parameters & MSE & Predicted orientation for the reference FMO \\\hline
    2-layer & $39607$ & $0.034$ & 
    $\{0.008, 0.016, 0.024, -0.037, 0.003, -0.027, -0.012 \}$ \\
    conv.\ & $6258$ & $0.062$ & 
    $\{0.0001, -0.037, 
  0.063, -0.064, -0.029, -0.002, -0.031\}$
    \\
    LSTM & $7087$ & $0.014$ & 
    $\{ 0.009, 0.005, 0.006, 0.004, 0.0,-0.002, -0.009\}$
    \end{tabular}
}
    \caption{Summary of the results obtained for the different NN algorithms (see ~\ref{sec:DIPOLE}) when trained $\{\mathbf{x:}$ 2DES $\rightarrow\mathbf{v:}$ Dipole Orientation$\}$. Second column indicates the number of fitting parameters underlying each of the algorithms. The mean squared error (MSE) of the trained algorithms is calculated in the test set not used during training. The predicted dipole moment for the FMO reference case is shown in the last column.}
    \label{tab:DIP}
\end{table*}
The LSTM algorithm, which includes a memory cell to simultaneously account for short and long term correlations between pixels, provides the best accuracy while keeping the number of underlying parameters small. We use this kind of NN algorithm in the next section.

\subsection{Prediction of noise averaged 2DES}

\begin{figure}[t]
(a) $\tau=100$~fs\\
\includegraphics[draft=false,width=0.99\textwidth]{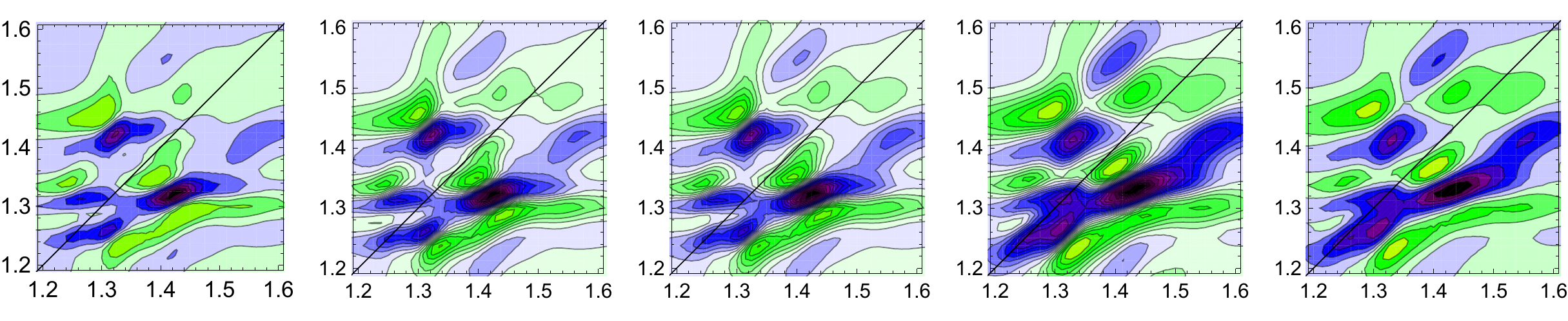}\\
(b)$\tau=500$~fs\\
\includegraphics[draft=false,width=0.99\textwidth]{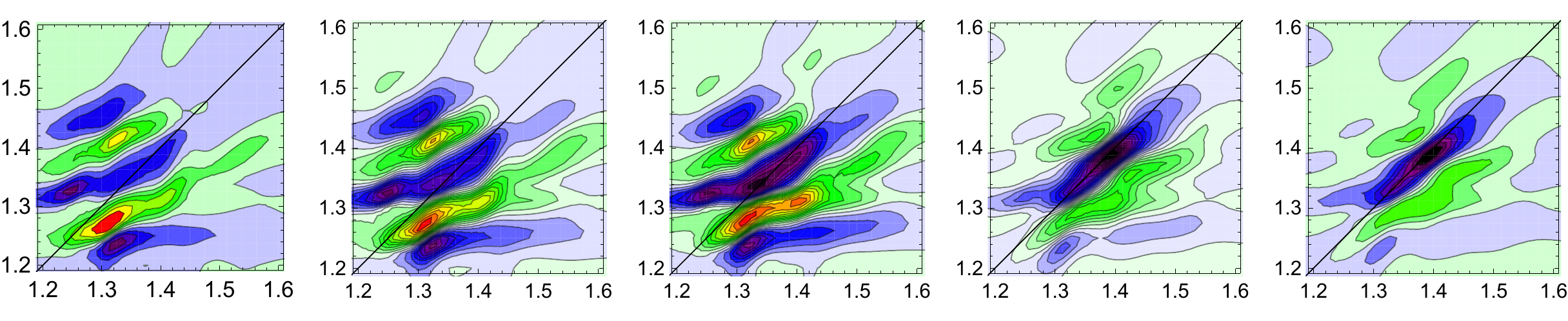}\\
(c)$\tau=1$~ps\\
\includegraphics[draft=false,width=0.99\textwidth]{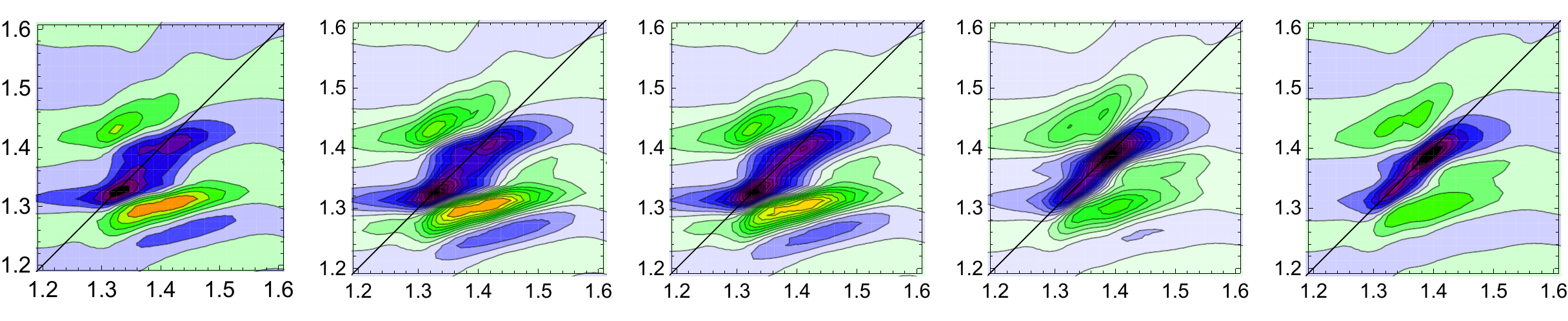}
(d)$\tau=1.8$~ps\\
\includegraphics[draft=false,width=0.99\textwidth]{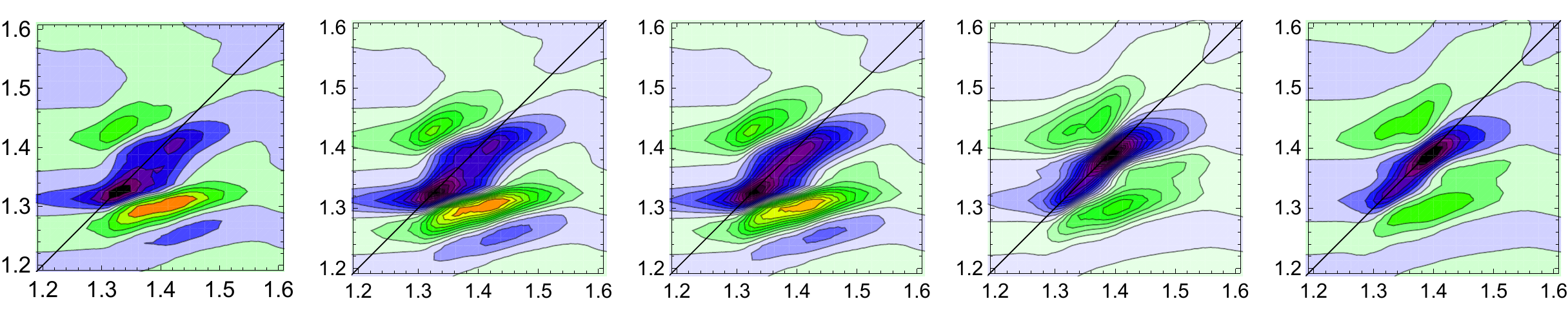}
\caption{Predicted averaged disordered FMO 2DE polarized rephasing spectra (axis in units of $1000$ cm$^-1$) for a laser polarisation sequence \cite{Thyrhaug2018} $\langle 45^\circ,-45^\circ,90^\circ,0^\circ \rangle$ at different delay times $\tau$ (rows) and different values of the disorder width (STD)(columns).
At each $\tau$ we use a LSTM algorithm trained over $4,500$ disorder realisations 
$\{\mathbf{x:} $  
disorder  
$\Delta \varepsilon \rightarrow \mathbf{v:}$ 
2DES $\}$ and predict the average spectra for $800$ disordered Hamiltonians of STD width.
Columns correspond to NN predictions for $\text{STD}=\{0,10,20,50\}$ and the left column shows the averaged training data.
}
\label{fig:FMOdis}
\end{figure}

The effect of disorder averaging in 2DES is still an open question. Our previous results \cite{Hein2012} showed that adding disorder for a linearly polarised sequence of pulses resulted in elongated blobs in the 2DES. We have found that for the specific polarisation sequence discussed in \cite{Thyrhaug2018}, noise averaging does more than just smearing out the blobs in the 2D spectra as shown in Fig.~\ref{fig:FMOdis}. 

\begin{figure}[htbp]
\centering
\includegraphics[draft=false,width=0.45\textwidth]{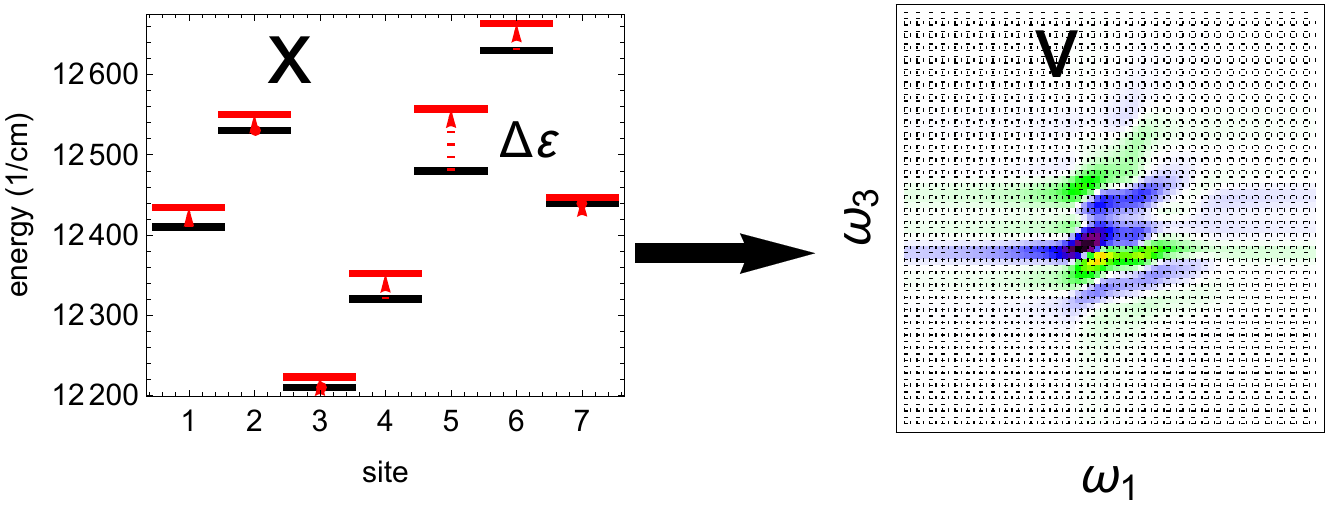}
\caption{We perform supervised learning experiments with $5,000$ data sets $\{\mathbf{x}\rightarrow\mathbf{v}\}$, where $\mathbf{x}$ is a $7$ dimensional real valued vector $\Delta\varepsilon$ that denotes the deviation of the FMO pigment site energies from the Adolphs-Renger Hamiltonian \cite{Adolphs2006a}. Data is generated randomly with a Gaussian distribution of width$=50$ cm$^{-1}$. Output images $\mathbf{v}$ are $65\times65$ pixel 2DES at several delay times $\tau$ calculated using HEOM for a polarised light sequence $\langle 45^\circ,-45^\circ,90^\circ,0^\circ\rangle$. Spectra calculations include the GB, SE and ESA rephasing pathways. }
\label{fig:TDIS}
\end{figure}

In order to systematically analyze the effect of disorder, we use a trained NN algorithm that generates 2DE rephasing spectra from input values for the static disorder of a FMO system (see \ref{sec:DISORDER} and Fig.~\ref{fig:TDIS}).  We find that a trained NN is an efficient way for storing and extraction 2DES disordered data. Results for $800$ disordered averaged FMO spectra at several delay times $\tau$ calculated using the trained NN algorithm are shown in Fig.~\ref{fig:FMOdis} for different widths of the static disorder. 

\section{Conclusions}
We have proven the capabilities of NN and supervised learning as an efficient tool for predicting parameters underlying 2D spectra.
In particular we have shown that using a trained NN one can predict the FMO dipole orientations with an accuracy of $0.01^\circ$, which is much better than current uncertainties in first principle calculations. We propose to use the trained NNs as an efficient way to encode computationally expensive theoretical models such as the HEOM used here to produce the data. 

As an example, the trained NN shown in \ref{tab:LSTMdipole} is a $\sim 30$~kB size file that contains the information of $9,000$  $28\times 28$ pixel sized 2DES computed with DM-HEOM, each spectra requiring about 1~h of compute time. We have also shown that we can use a trained NN to produce 2DE spectra from related parameters such as the pigment site energies. This results in an interesting application of using trained NN algorithms to efficiently calculate disordered averaged spectra, which is very demanding computationally and usually avoided or very roughly approximated.
We showcase the use of trained NN to analyse the non-trivial effect of disorder averaging for polarization controlled 2DES and systematically calculate disordered spectra for increasing disorder widths. We have compared simple multilayer, convolutional and recurrent neural networks. We observe better accuracy and smaller number of underlying parameters with recurrent LSTM algorithms. Training the algorithms requires modest data set sizes and running times, indicating that feature extraction from spectra images can be efficiently achieved with NN algorithms. Moreover, we have shown that one can predict spectra images from a small set of parameters demonstrating that the correlation of system parameters and spectra images is efficiently encoded in a NN type algorithm. 

Having proven the capabilities of NN and machine learning with computationally produced data, it would be interesting to explore the capabilities of these methods when trained with experimental data or computation/experimental mixed data sets. 

\section*{Acknowledgements}
This contribution is dedicated to Leonas Valkunas, whose contribution to excitonic energy transfer in molecules advanced the understanding of photosynthetic processes.
The work was supported by the North-German Supercomputing Alliance (HLRN) and by the German Research Foundation (DFG), grants KR~2889 and RE~1389.
M.R. has received funding from the European Union's Horizon 2020 research and innovation programme under the Marie Sklodowska-Curie grant agreement No 707636.

\clearpage

\appendix

\section{Networks}
\label{sec:AppendixNN}
\subsection{Dipole orientations}
\label{sec:DIPOLE}
We train different NN algorithms by assigning 2DES to dipole orientation vectors $\{\mathbf{x}:$ 2DES $\rightarrow \mathbf{v}: \Delta \alpha\}_l$ (see Fig.~\ref{fig:TD}) for $l=1,\ldots,N$  where $N=9,000$. 
The $\Delta \alpha_l$ are generated using a rectangular random number generator between $\pm 10$.
Final network configurations used are summarised in Tables \ref{tab:2NNdipole}, \ref{tab:CONVdipole} and \ref{tab:LSTMdipole}. The accuracy (MSE) obtained with each network and for the FMO reference case are summarized in Table \ref{tab:DIP}.
\subsubsection{Single hidden layer}
\label{sec:2NN}
We use a simple $2$-layer neural network algorithm in Eq \ref{eq:NN} for $M=2$, as described in Eq.~(\ref{eq:2N}). 
To find the optimal network configuration \cite{Lawrence1996} we vary the parameters in the training, including the number of hidden layers $M$, the dimension of a single hidden layer $N_h$, the number of training data $N$, the number of epochs and the batch size $B$. We did not find significant improvements in the accuracy when increasing the number of layers in the network $M$ nor increasing the number of training data $N$. We calculated the MSE as a function of the dimension of the hidden layer $N_h$ and observe slight decrease with increasing $N_h$.  Increasing $N_h$ implies a linear increase of the number of fitting parameters and requires higher number of epochs to achieve convergence. 
Table \ref{tab:2NNdipole} shows a sketch of the NN algorithm and the number of fitting parameters $ \mathbf{b}^1,\mathbf{W}^1, \mathbf{b}^2,\mathbf{W}^2$. Training the algorithm takes 2.9~h on a NVIDIA GeForce GTX 1080 Ti with 11 GB GDDR5X-RAM.
\begin{table*}
\begin{tabular}{c|c}
\parbox{0.5\textwidth}{\includegraphics[width=0.49\textwidth]{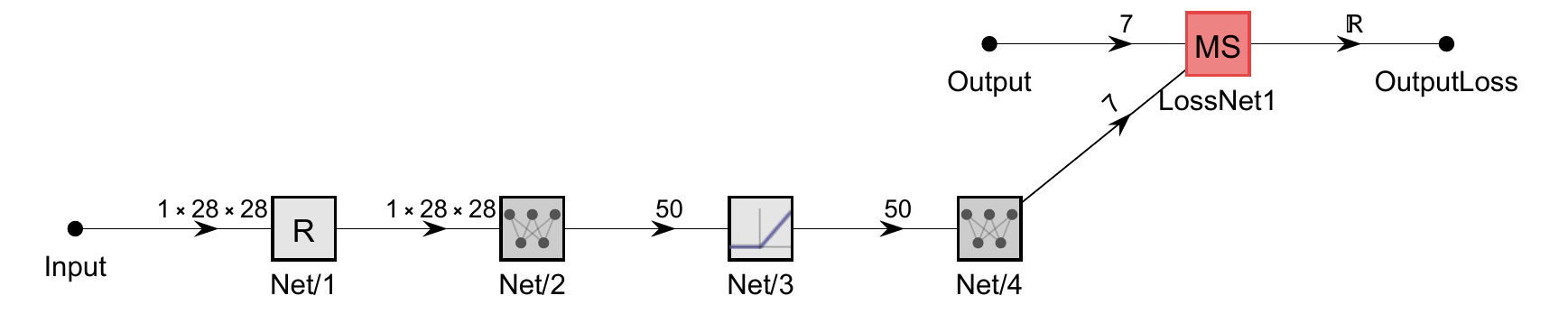}\\
\includegraphics[width=0.4\textwidth]{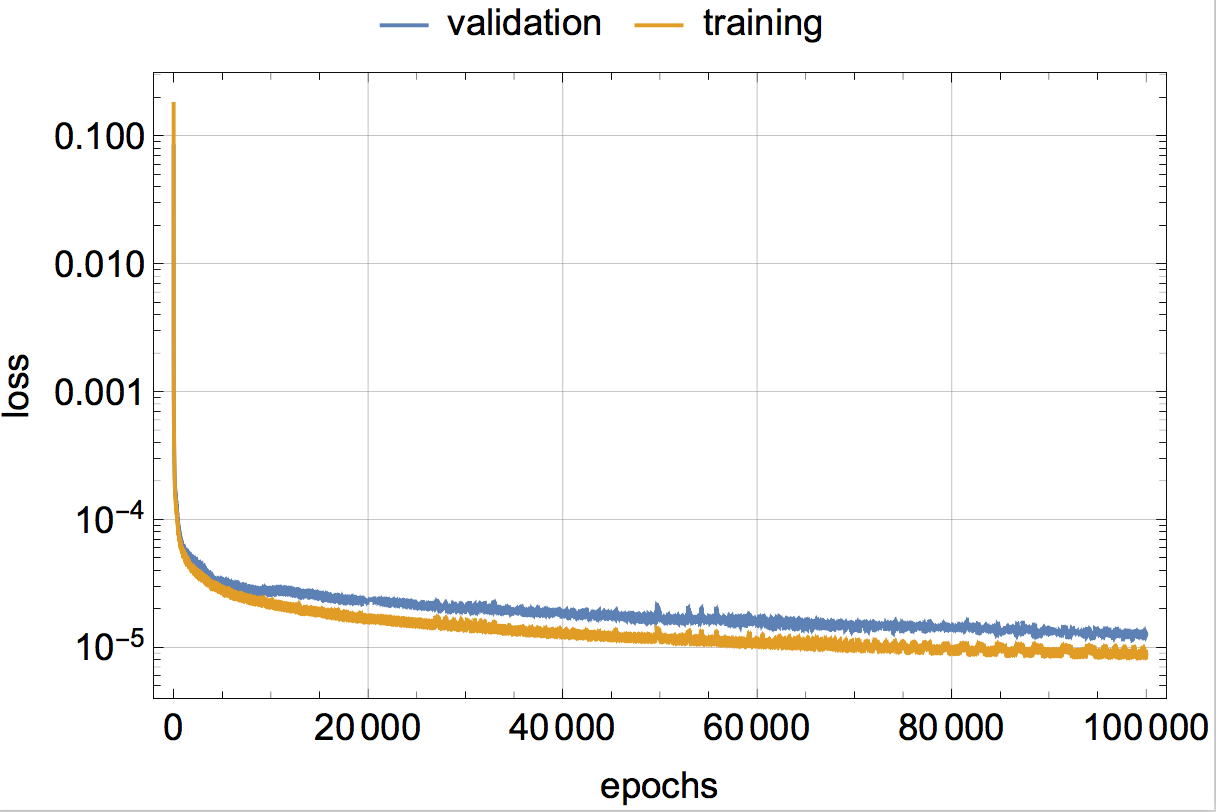}}     & 
$
\begin{array}{ccc}
\text{layer} & \text{param type} & \text{\# param} \\ 
 2 & \text{Biases} & 50 \\
 2 & \text{Weights} & 39200 \\
 4 & \text{Biases} & 7 \\
 4 & \text{Weights} & 350 \\
\end{array}
$
\end{tabular}
\caption{2-NN used for predicting dipole orientation from 2DES $\{ \mathbf{x:}$ 2DES $\rightarrow$  $ \mathbf{v:}$ dipole orientation$\}$. Up-left: input to output network diagram. Blocks in the diagram include R: reshape linear matrix, W: linear operation and $\_$/: ramp function. Down-left: Loss function Eq.~\ref{eq:loss} as a function of the training rounds (epochs) calculated over the training and validation set. Loss function for the small validation set not seen by the algorithm is above the training loss function indicating that there is no overfitting. Right: number of fitting parameters of the two active layers in the NN.}
\label{tab:2NNdipole}
\end{table*}

\subsubsection{Convolutional Neural Network}
Convolutional NN alternate convolutional layers that apply a filter (called kernel) into the input matrix with pooling layers that perform subsampling \cite{LeCun}. Looking for correlations among the spectral peaks, we use kernels of dimension image size over the number of pigments ($28/7$).  In order to perform meaningful comparisons with the different algorithms we use similar $B$ and $N$ as in the previous section.
Table~\ref{tab:CONVdipole} shows a sketch of the NN algorithm and the number of fitting parameters. Training the algorithm takes 3.3~h on a NVIDIA GeForce GTX 1080 Ti with 11 GB GDDR5X-RAM.
\begin{table*}
\begin{tabular}{c|c}
\parbox{0.5\textwidth}{\includegraphics[width=0.49\textwidth]{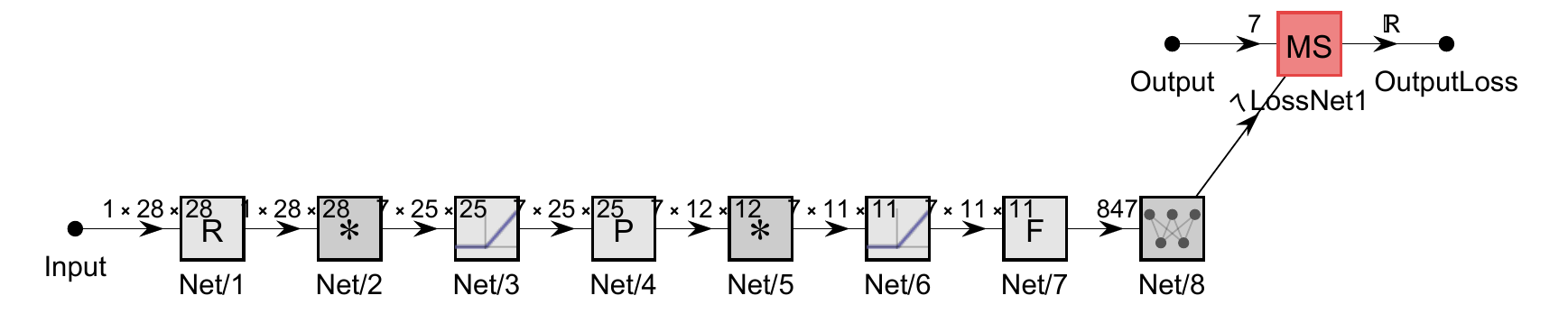}\\
\includegraphics[width=0.4\textwidth]{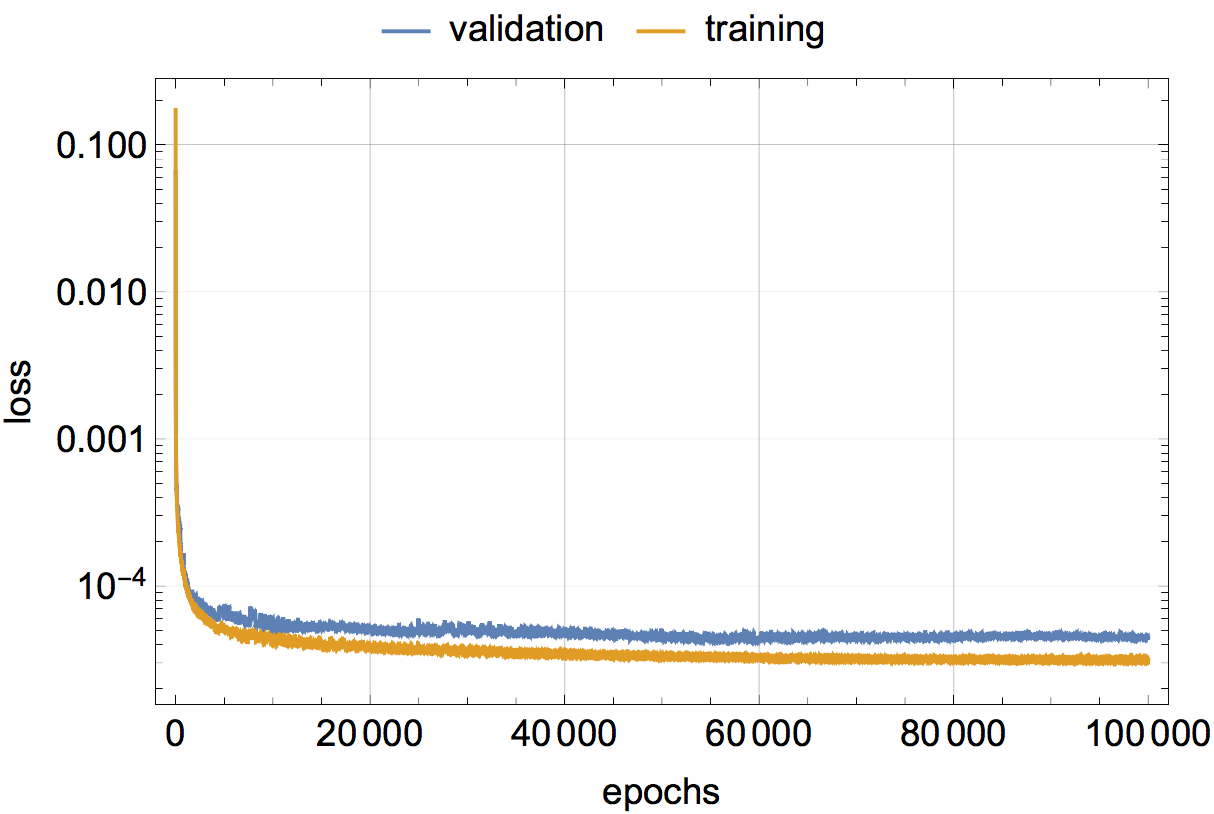}}     & 
$
\begin{array}{ccc}
 \text{layer} & \text{param type} & \text{\# param} \\
 2 & \text{Biases} & 7 \\
 2 & \text{Weights} & 112 \\
 5 & \text{Biases} & 7 \\
 5 & \text{Weights} & 196 \\
 8 & \text{Biases} & 7 \\
 8 & \text{Weights} & 5929 \\
\end{array}
$
\end{tabular}
\caption{Convolution NN used for prediction of the dipole orientation from input 2DES.
Up-left: input to output network diagram. Blocks in the diagram include R: reshape linear matrix, $*$: convolution layer, $\_$/: ramp function, P: pooling layer, F: flatten layer and W: linear layer (see Mathematica Documentation in \cite{Wolfram2018}). Down-left: Loss function Eq. (\ref{eq:loss}) as a function of the training rounds (epochs) calculated over the training and validation set. Right: number of fitting parameters of the NN  (convolution and linear layers)}
\label{tab:CONVdipole}
\end{table*}

\subsubsection{LSTM}
Long Short Term Memory (LSTM) recurrent neural networks contain a non-linear recurrent cell with forget and memory layers which allows for retaining information across the data in a non trivial way\cite{Hochreiter1997}. The number of fitting parameters increases linearly with the dimension of the output of the LSTM layer. 
Even with small output dimension the LSTM outperforms the other NN tested here. 
Table~\ref{tab:LSTMdipole} shows a sketch of the NN algorithm and the number of fitting parameters. Training the algorithm takes 7.1~h on a NVIDIA GeForce GTX 1080 Ti with 11 GB GDDR5X-RAM.
\begin{table*}
\begin{tabular}{c|c}
\parbox{0.4\textwidth}{\includegraphics[width=0.49\textwidth]{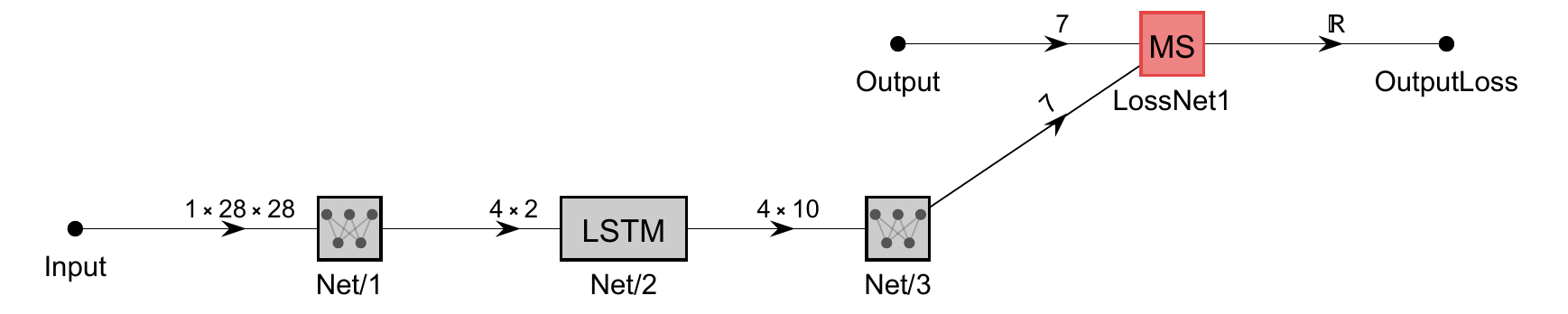}\\
\includegraphics[width=0.4\textwidth]{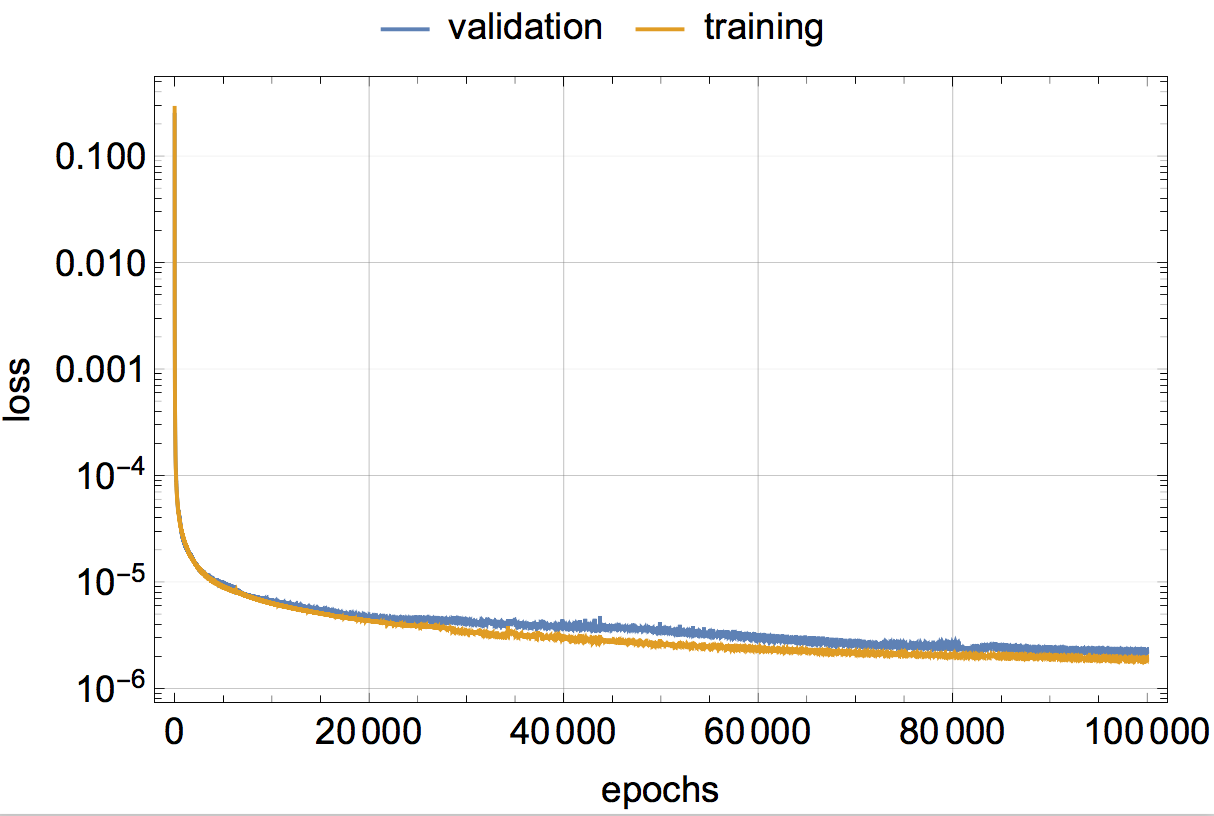}}     & 
{\small
$
\begin{array}{ccc}
\text{layer} & \text{param type} & \text{\# param} \\ 
 1 & \text{Biases} & 8 \\
 1 & \text{Weights} & 6272 \\
 2 & \text{ForgetGateBiases} & 10 \\
 2 & \text{ForgetGateInputWeights} & 20 \\
 2 & \text{ForgetGateStateWeights} & 100 \\
 2 & \text{InputGateBiases} & 10 \\
 2 & \text{InputGateInputWeights} & 20 \\
 2 & \text{InputGateStateWeights} & 100 \\
 2 & \text{MemoryGateBiases} & 10 \\
 2 & \text{MemoryGateInputWeights} & 20 \\
 2 & \text{MemoryGateStateWeights} & 100 \\
 2 & \text{OutputGateBiases} & 10 \\
 2 & \text{OutputGateInputWeights} & 20 \\
 2 & \text{OutputGateStateWeights} & 100 \\
 3 & \text{Biases} & 7 \\
 3 & \text{Weights} & 280 \\
\end{array}
$
}
\end{tabular}
\caption{Recurrent NN used for prediction of the dipole orientation from input 2DES.
Up-left: input to output network diagram. Blocks in the diagram include W: linear layer, LSTM: Long Short Term Memory layer (see Mathematica Documentation in \cite{Wolfram2018}). Down-left: Loss function Eq.~(\ref{eq:loss}) as a function of the training rounds (epochs) calculated over the training and validation set. Right: number of fitting parameters of the NN including the input, forget, memory and output gates comprising the LSTM algorithm \cite{Hochreiter1997}.}
\label{tab:LSTMdipole}
\end{table*}
\subsection{Disordered Spectra}
\label{sec:DISORDER}
We train the NN by assigning the $7$ values of the static disorder to the resulting 2DES $\{\mathbf{x}: \Delta \varepsilon \rightarrow \mathbf{v}:$ 2DES $\}_l$ for $l=1,.,N$ as shown in Fig.~\ref{fig:TDIS}. 
To generate the training data we calculate the rephasing 2D electronic spectra at fixed delay times $\tau=100$,$500$,$1000$ and $1800$ fs for the polarisation sequence $\langle 45^\circ,-45^\circ,90^\circ,0^\circ\rangle$ \cite{Thyrhaug2018} for $5,000$ random deviations of the Adolphs-Renger FMO site energies\cite{Adolphs2006a}. We generate $\Delta \varepsilon_l$ with a gaussian distribution of width STD$=50$cm$^{-1}$.

We normalise individual spectra $\mathrm{x}_l$ by a common factor (the maximum of all training points) and the disorder vector $\mathrm{v}_l$ is normalised to $200$ cm$^{-1}$. We train the LSTM NN algorithm described in Table \ref{tab:LSTMdisorder}. This takes 2.9~h for the $\tau=1.8$~ps 2DES data on a NVIDIA GeForce GTX 1080 Ti with 11 GB GDDR5X-RAM. We fix the number of epochs to $100,000$ and the batch size to $100$.

\begin{table*}
\begin{tabular}{c|c}
\parbox{0.4\textwidth}{\includegraphics[width=0.4\textwidth]{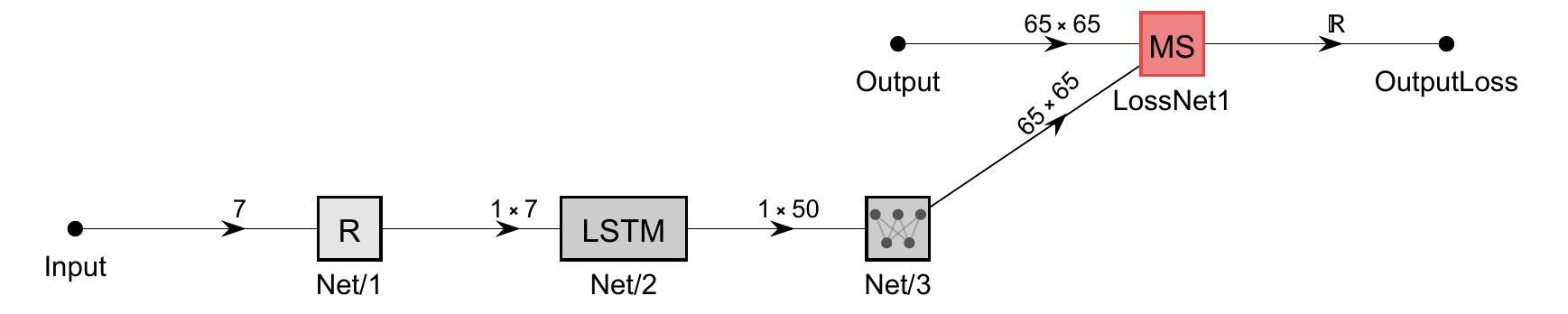}\\
\includegraphics[width=0.4\textwidth]{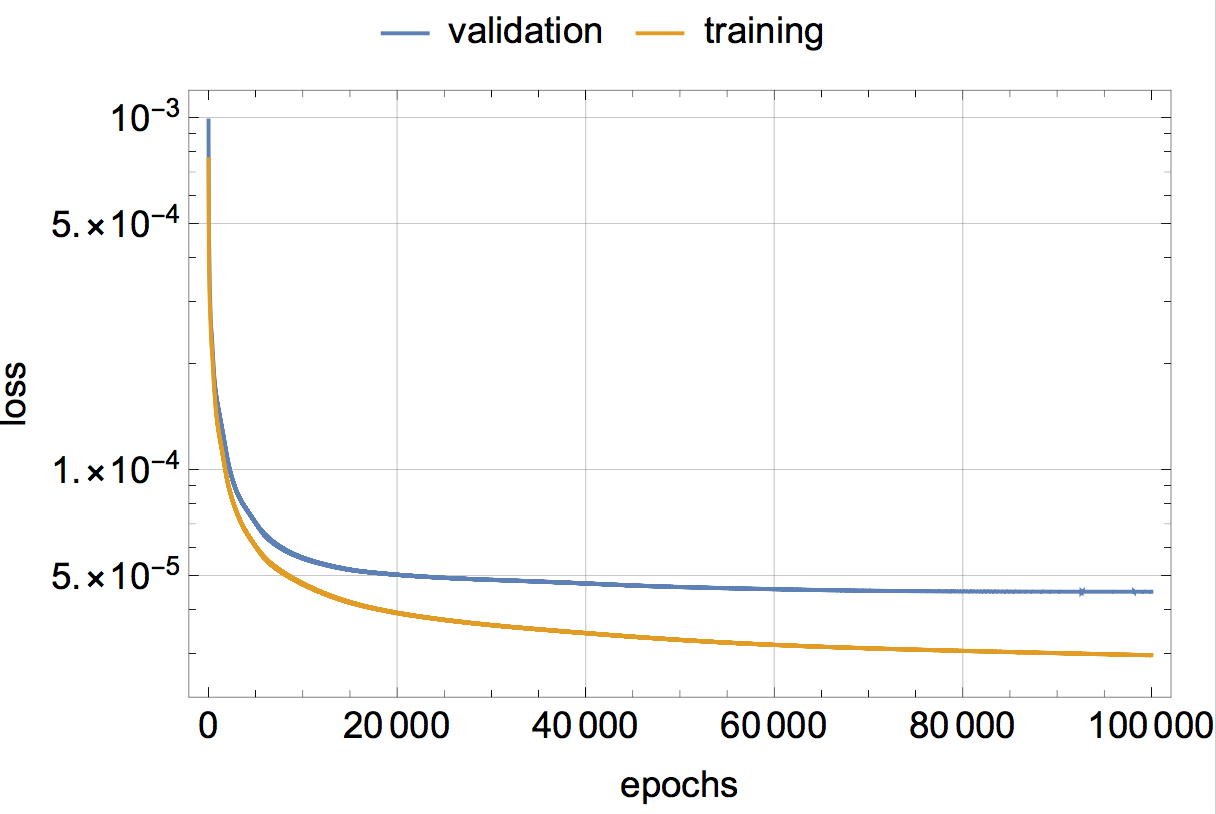}}     & 
{\small 
$
\begin{array}{ccc}
\text{layer} & \text{param type} & \text{\# param} \\ 
 2 & \text{ForgetGateBiases} & 50 \\
 2 & \text{ForgetGateInputWeights} & 350 \\
 2 & \text{ForgetGateStateWeights} & 2500 \\
 2 & \text{InputGateBiases} & 50 \\
 2 & \text{InputGateInputWeights} & 350 \\
 2 & \text{InputGateStateWeights} & 2500 \\
 2 & \text{MemoryGateBiases} & 50 \\
 2 & \text{MemoryGateInputWeights} & 350 \\
 2 & \text{MemoryGateStateWeights} & 2500 \\
 2 & \text{OutputGateBiases} & 50 \\
 2 & \text{OutputGateInputWeights} & 350 \\
 2 & \text{OutputGateStateWeights} & 2500 \\
 3 & \text{Biases} & 4225 \\
 3 & \text{Weights} & 211250 \\
\end{array}
$
}
\end{tabular}
\caption{Recurrent (LSTM) algorithm used for predicting 2DES from disordered FMO Hamiltonians $\{ \mathbf{x:}$ site energies $\rightarrow$  $ \mathbf{v:}$ 2DES $\}$.
Up-left: input to output network diagram. Blocks in the diagram include R: reshape layer, LSTM: Long Short Term Memory layer, W: linear layer (see Mathematica Documentation in \cite{Wolfram2018}). Down-left: Loss function Eq. (\ref{eq:loss}) as a function of the training rounds (epochs) calculated over the training and validation set. Right: number of fitting parameters of the NN including the input, forget, memory and output gates comprising the LSTM algorithm \cite{Hochreiter1997} and the linear layer.
}
\label{tab:LSTMdisorder}
\end{table*}


\begin{thebibliography}{46}
\expandafter\ifx\csname natexlab\endcsname\relax\def\natexlab#1{#1}\fi
\providecommand{\bibinfo}[2]{#2}
\ifx\xfnm\relax \def\xfnm[#1]{\unskip,\space#1}\fi
\bibitem[{Adolphs and Renger(2006)}]{Adolphs2006a}
\bibinfo{author}{J.~Adolphs}, \bibinfo{author}{T.~Renger},
\newblock \bibinfo{title}{{How Proteins Trigger Excitation Energy Transfer in
  the FMO Complex of Green Sulfur Bacteria}},
\newblock \bibinfo{journal}{Biophysical Journal} \bibinfo{volume}{91}
  (\bibinfo{year}{2006}) \bibinfo{pages}{2778--2797}.
\bibitem[{Sanchez-Gonzalez et~al.(2017)Sanchez-Gonzalez, Micaelli, Olivier,
  Barillot, Ilchen, Lutman, Marinelli, Maxwell, Achner, Ag{\aa}ker, Berrah,
  Bostedt, Bozek, Buck, Bucksbaum, Montero, Cooper, Cryan, Dong, Feifel,
  Frasinski, Fukuzawa, Galler, Hartmann, Hartmann, Helml, Johnson, Knie,
  Lindahl, Liu, Motomura, Mucke, O'Grady, Rubensson, Simpson, Squibb,
  S{\aa}the, Ueda, Vacher, Walke, Zhaunerchyk, Coffee, and
  Marangos}]{Micaelli2017}
\bibinfo{author}{A.~Sanchez-Gonzalez}, \bibinfo{author}{P.~Micaelli},
  \bibinfo{author}{C.~Olivier}, \bibinfo{author}{T.~R. Barillot},
  \bibinfo{author}{M.~Ilchen}, \bibinfo{author}{A.~A. Lutman},
  \bibinfo{author}{A.~Marinelli}, \bibinfo{author}{T.~Maxwell},
  \bibinfo{author}{A.~Achner}, \bibinfo{author}{M.~Ag{\aa}ker},
  \bibinfo{author}{N.~Berrah}, \bibinfo{author}{C.~Bostedt},
  \bibinfo{author}{J.~D. Bozek}, \bibinfo{author}{J.~Buck},
  \bibinfo{author}{P.~H. Bucksbaum}, \bibinfo{author}{S.~C. Montero},
  \bibinfo{author}{B.~Cooper}, \bibinfo{author}{J.~P. Cryan},
  \bibinfo{author}{M.~Dong}, \bibinfo{author}{R.~Feifel},
  \bibinfo{author}{L.~J. Frasinski}, \bibinfo{author}{H.~Fukuzawa},
  \bibinfo{author}{A.~Galler}, \bibinfo{author}{G.~Hartmann},
  \bibinfo{author}{N.~Hartmann}, \bibinfo{author}{W.~Helml},
  \bibinfo{author}{A.~S. Johnson}, \bibinfo{author}{A.~Knie},
  \bibinfo{author}{A.~O. Lindahl}, \bibinfo{author}{J.~Liu},
  \bibinfo{author}{K.~Motomura}, \bibinfo{author}{M.~Mucke},
  \bibinfo{author}{C.~O'Grady}, \bibinfo{author}{J.-E. Rubensson},
  \bibinfo{author}{E.~R. Simpson}, \bibinfo{author}{R.~J. Squibb},
  \bibinfo{author}{C.~S{\aa}the}, \bibinfo{author}{K.~Ueda},
  \bibinfo{author}{M.~Vacher}, \bibinfo{author}{D.~J. Walke},
  \bibinfo{author}{V.~Zhaunerchyk}, \bibinfo{author}{R.~N. Coffee},
  \bibinfo{author}{J.~P. Marangos},
\newblock \bibinfo{title}{{Accurate prediction of X-ray pulse properties from a
  free-electron laser using machine learning}},
\newblock \bibinfo{journal}{Nature Communications} \bibinfo{volume}{8}
  (\bibinfo{year}{2017}) \bibinfo{pages}{15461}.
\bibitem[{Roeding et~al.(2017)Roeding, Klimovich, and Brixner}]{Roeding2017a}
\bibinfo{author}{S.~Roeding}, \bibinfo{author}{N.~Klimovich},
  \bibinfo{author}{T.~Brixner},
\newblock \bibinfo{title}{{Optimizing sparse sampling for 2D electronic
  spectroscopy}},
\newblock \bibinfo{journal}{The Journal of Chemical Physics}
  \bibinfo{volume}{146} (\bibinfo{year}{2017}) \bibinfo{pages}{084201}.
\bibitem[{Blankenship(2014)}]{Blankenship2014}
\bibinfo{author}{R.~E. Blankenship}, \bibinfo{title}{{Molecular Mechanisms of
  Photosynthesis}}, \bibinfo{publisher}{Wiley}, \bibinfo{address}{Oxford, UK},
  \bibinfo{edition}{2nd} edition, \bibinfo{year}{2014}.
\bibitem[{Chmeliov et~al.(2016)Chmeliov, Gelzinis, Songaila, Augulis, Duffy,
  Ruban, and Valkunas}]{Chmeliov2016}
\bibinfo{author}{J.~Chmeliov}, \bibinfo{author}{A.~Gelzinis},
  \bibinfo{author}{E.~Songaila}, \bibinfo{author}{R.~Augulis},
  \bibinfo{author}{C.~D.~P. Duffy}, \bibinfo{author}{A.~V. Ruban},
  \bibinfo{author}{L.~Valkunas},
\newblock \bibinfo{title}{{The nature of self-regulation in photosynthetic
  light-harvesting antenna}},
\newblock \bibinfo{journal}{Nature Plants} \bibinfo{volume}{2}
  (\bibinfo{year}{2016}) \bibinfo{pages}{16045}.
\bibitem[{Tanimura and Kubo(1989)}]{Tanimura1989}
\bibinfo{author}{Y.~Tanimura}, \bibinfo{author}{R.~Kubo},
\newblock \bibinfo{title}{{Time Evoultion of a Quantum System in Contact with a
  Nearly Gussian-Markoffian Noise Bath}},
\newblock \bibinfo{journal}{Journal of the Physics Society Japan}
  \bibinfo{volume}{58} (\bibinfo{year}{1989}) \bibinfo{pages}{101--114}.
\bibitem[{Ishizaki and Fleming(2009)}]{Ishizaki2009}
\bibinfo{author}{A.~Ishizaki}, \bibinfo{author}{G.~R. Fleming},
\newblock \bibinfo{title}{{Unified treatment of quantum coherent and incoherent
  hopping dynamics in electronic energy transfer: Reduced hierarchy equation
  approach}},
\newblock \bibinfo{journal}{The Journal of Chemical Physics}
  \bibinfo{volume}{130} (\bibinfo{year}{2009}) \bibinfo{pages}{234111}.
\bibitem[{Kreisbeck et~al.(2011)Kreisbeck, Kramer, Rodr{\'{i}}guez, and
  Hein}]{Kreisbeck2011}
\bibinfo{author}{C.~Kreisbeck}, \bibinfo{author}{T.~Kramer},
  \bibinfo{author}{M.~Rodr{\'{i}}guez}, \bibinfo{author}{B.~Hein},
\newblock \bibinfo{title}{{High-performance solution of hierarchical equations
  of motion for studying energy transfer in light-harvesting complexes}},
\newblock \bibinfo{journal}{Journal of Chemical Theory and Computation}
  \bibinfo{volume}{7} (\bibinfo{year}{2011}) \bibinfo{pages}{2166--2174}.
\bibitem[{Hein et~al.(2012)Hein, Kreisbeck, Kramer, and
  Rodr{\'{i}}guez}]{Hein2012}
\bibinfo{author}{B.~Hein}, \bibinfo{author}{C.~Kreisbeck},
  \bibinfo{author}{T.~Kramer}, \bibinfo{author}{M.~Rodr{\'{i}}guez},
\newblock \bibinfo{title}{{Modelling of oscillations in two-dimensional
  echo-spectra of the Fenna-Matthews-Olson complex}},
\newblock \bibinfo{journal}{New Journal of Physics} \bibinfo{volume}{14}
  (\bibinfo{year}{2012}) \bibinfo{pages}{023018}.
\bibitem[{Somsen et~al.(1996)Somsen, van Grondelle, and van
  Amerongen}]{Somsen1996}
\bibinfo{author}{O.~Somsen}, \bibinfo{author}{R.~van Grondelle},
  \bibinfo{author}{H.~van Amerongen},
\newblock \bibinfo{title}{{Spectral broadening of interacting pigments:
  polarized absorption by photosynthetic proteins}},
\newblock \bibinfo{journal}{Biophysical Journal} \bibinfo{volume}{71}
  (\bibinfo{year}{1996}) \bibinfo{pages}{1934--1951}.
\bibitem[{LeCun and Bengio(1988)}]{Lecun1998}
\bibinfo{author}{Y.~LeCun}, \bibinfo{author}{Y.~Bengio},
  \bibinfo{title}{{Convolutional networks for images, speech, and time
  series}}, \bibinfo{edition}{the handbo} edition, \bibinfo{year}{1988}.
\bibitem[{Hinton et~al.(2006)Hinton, Osindero, and Teh}]{Hinton2006}
\bibinfo{author}{G.~E. Hinton}, \bibinfo{author}{S.~Osindero},
  \bibinfo{author}{Y.-W. Teh},
\newblock \bibinfo{title}{{A Fast Learning Algorithm for Deep Belief Nets}},
\newblock \bibinfo{journal}{Neural Computation} \bibinfo{volume}{18}
  (\bibinfo{year}{2006}) \bibinfo{pages}{1527--1554}.
\bibitem[{Cireşan et~al.(2011)Cireşan, Meier, Masci, Gambardella, and
  Schmidhuber}]{Krizhevsky2012}
\bibinfo{author}{D.~C. Cireşan}, \bibinfo{author}{U.~Meier},
  \bibinfo{author}{J.~Masci}, \bibinfo{author}{L.~M. Gambardella},
  \bibinfo{author}{J.~Schmidhuber},
\newblock \bibinfo{title}{{High-Performance Neural Networks for Visual Object
  Classification}},
\newblock \bibinfo{journal}{Lancet (London, England)} \bibinfo{volume}{346}
  (\bibinfo{year}{2011}) \bibinfo{pages}{1501}.
\bibitem[{Goodfellow et~al.(2016)Goodfellow, Bengio, and
  Courville}]{Goodfellow2016}
\bibinfo{author}{I.~Goodfellow}, \bibinfo{author}{Y.~Bengio},
  \bibinfo{author}{A.~Courville}, \bibinfo{title}{{Deep Learning}},
  \bibinfo{publisher}{MIT Press}, \bibinfo{year}{2016}.
\bibitem[{van Nieuwenburg et~al.(2017)van Nieuwenburg, Liu, and
  Huber}]{VanNieuwenburg2016}
\bibinfo{author}{E.~P.~L. van Nieuwenburg}, \bibinfo{author}{Y.-H. Liu},
  \bibinfo{author}{S.~D. Huber},
\newblock \bibinfo{title}{{Learning phase transitions by confusion}},
\newblock \bibinfo{journal}{Nature Physics}  (\bibinfo{year}{2017})
  \bibinfo{pages}{5}.
\bibitem[{Carrasquilla and Melko(2017)}]{Carrasquilla2016}
\bibinfo{author}{J.~Carrasquilla}, \bibinfo{author}{R.~G. Melko},
\newblock \bibinfo{title}{{Machine learning phases of matter}},
\newblock \bibinfo{journal}{Nature Physics}  (\bibinfo{year}{2017})
  \bibinfo{pages}{1--18}.
\bibitem[{Mavadia et~al.(2017)Mavadia, Frey, Sastrawan, Dona, and
  Biercuk}]{Mavadia2016}
\bibinfo{author}{S.~Mavadia}, \bibinfo{author}{V.~Frey},
  \bibinfo{author}{J.~Sastrawan}, \bibinfo{author}{S.~Dona},
  \bibinfo{author}{M.~J. Biercuk},
\newblock \bibinfo{title}{{Prediction and real-time compensation of qubit
  decoherence via machine learning}},
\newblock \bibinfo{journal}{Nature Communications} \bibinfo{volume}{8}
  (\bibinfo{year}{2017}) \bibinfo{pages}{14106}.
\bibitem[{Carleo and Troyer(2017)}]{Carleo2017}
\bibinfo{author}{G.~Carleo}, \bibinfo{author}{M.~Troyer},
\newblock \bibinfo{title}{{Solving the quantum many-body problem with
  artificial neural networks}},
\newblock \bibinfo{journal}{Science} \bibinfo{volume}{355}
  (\bibinfo{year}{2017}) \bibinfo{pages}{602--606}.
\bibitem[{H{\"{a}}se et~al.(2017)H{\"{a}}se, Kreisbeck, and
  Aspuru-Guzik}]{Hase2017}
\bibinfo{author}{F.~H{\"{a}}se}, \bibinfo{author}{C.~Kreisbeck},
  \bibinfo{author}{A.~Aspuru-Guzik},
\newblock \bibinfo{title}{{Machine learning for quantum dynamics: deep learning
  of excitation energy transfer properties}},
\newblock \bibinfo{journal}{Chemical Science} \bibinfo{volume}{8}
  (\bibinfo{year}{2017}) \bibinfo{pages}{8419--8426}.
\bibitem[{Bandyopadhyay et~al.(2018)Bandyopadhyay, Huang, Sun, and
  Zhao}]{Bandyopadhyay2018}
\bibinfo{author}{S.~Bandyopadhyay}, \bibinfo{author}{Z.~Huang},
  \bibinfo{author}{K.~Sun}, \bibinfo{author}{Y.~Zhao},
\newblock \bibinfo{title}{{Applications of neural networks to the simulation of
  dynamics of open quantum systems}},
\newblock \bibinfo{journal}{Chemical Physics}  (\bibinfo{year}{2018}).
\bibitem[{Dral et~al.(2018)Dral, Barbatti, and Thiel}]{Dral2018}
\bibinfo{author}{P.~O. Dral}, \bibinfo{author}{M.~Barbatti},
  \bibinfo{author}{W.~Thiel},
\newblock \bibinfo{title}{{Nonadiabatic Excited-State Dynamics with Machine
  Learning}},
\newblock \bibinfo{journal}{The Journal of Physical Chemistry Letters}
  (\bibinfo{year}{2018}) \bibinfo{pages}{5660--5663}.
\bibitem[{Rupp et~al.(2012)Rupp, Tkatchenko, M{\"{u}}ller, and von
  Lilienfeld}]{Rupp2012}
\bibinfo{author}{M.~Rupp}, \bibinfo{author}{A.~Tkatchenko},
  \bibinfo{author}{K.-R. M{\"{u}}ller}, \bibinfo{author}{O.~A. von Lilienfeld},
\newblock \bibinfo{title}{{Fast and Accurate Modeling of Molecular Atomization
  Energies with Machine Learning}},
\newblock \bibinfo{journal}{Physical Review Letters} \bibinfo{volume}{108}
  (\bibinfo{year}{2012}) \bibinfo{pages}{058301}.
\bibitem[{Montavon et~al.(2013)Montavon, Rupp, Gobre, Vazquez-Mayagoitia,
  Hansen, Tkatchenko, M{\"{u}}ller, and {Anatole Von
  Lilienfeld}}]{Montavon2013}
\bibinfo{author}{G.~Montavon}, \bibinfo{author}{M.~Rupp},
  \bibinfo{author}{V.~Gobre}, \bibinfo{author}{A.~Vazquez-Mayagoitia},
  \bibinfo{author}{K.~Hansen}, \bibinfo{author}{A.~Tkatchenko},
  \bibinfo{author}{K.~R. M{\"{u}}ller}, \bibinfo{author}{O.~{Anatole Von
  Lilienfeld}},
\newblock \bibinfo{title}{{Machine learning of molecular electronic properties
  in chemical compound space}},
\newblock \bibinfo{journal}{New Journal of Physics} \bibinfo{volume}{15}
  (\bibinfo{year}{2013}).
\bibitem[{Johnson and Karanicolas(2016)}]{Duvenaud2015}
\bibinfo{author}{D.~K. Johnson}, \bibinfo{author}{J.~Karanicolas},
\newblock \bibinfo{title}{{Ultra-High-Throughput Structure-Based Virtual
  Screening for Small-Molecule Inhibitors of Protein–Protein Interactions}},
\newblock \bibinfo{journal}{Journal of Chemical Information and Modeling}
  \bibinfo{volume}{56} (\bibinfo{year}{2016}) \bibinfo{pages}{399--411}.
\bibitem[{Rupp(2015)}]{Rupp2015a}
\bibinfo{author}{M.~Rupp},
\newblock \bibinfo{title}{{Special issue on machine learning and quantum
  mechanics}},
\newblock \bibinfo{journal}{International Journal of Quantum Chemistry}
  \bibinfo{volume}{115} (\bibinfo{year}{2015}) \bibinfo{pages}{1003--1004}.
\bibitem[{Thyrhaug et~al.(2018)Thyrhaug, Tempelaar, Alcocer, {\v{Z}}{\'{i}}dek,
  B{\'{i}}na, Knoester, Jansen, and Zigmantas}]{Thyrhaug2018}
\bibinfo{author}{E.~Thyrhaug}, \bibinfo{author}{R.~Tempelaar},
  \bibinfo{author}{M.~J. Alcocer}, \bibinfo{author}{K.~{\v{Z}}{\'{i}}dek},
  \bibinfo{author}{D.~B{\'{i}}na}, \bibinfo{author}{J.~Knoester},
  \bibinfo{author}{T.~L. Jansen}, \bibinfo{author}{D.~Zigmantas},
\newblock \bibinfo{title}{{Identification and characterization of diverse
  coherences in the Fenna–Matthews–Olson complex}},
\newblock \bibinfo{journal}{Nature Chemistry} \bibinfo{volume}{10}
  (\bibinfo{year}{2018}) \bibinfo{pages}{1--7}.
\bibitem[{Noack et~al.(2018)Noack, Reinefeld, Kramer, and Steinke}]{Noack2018}
\bibinfo{author}{M.~Noack}, \bibinfo{author}{A.~Reinefeld},
  \bibinfo{author}{T.~Kramer}, \bibinfo{author}{T.~Steinke},
\newblock \bibinfo{title}{{DM-HEOM: A Portable and Scalable Solver-Framework
  for the Hierarchical Equations of Motion}},
\newblock in: \bibinfo{booktitle}{2018 IEEE International Parallel and
  Distributed Processing Symposium Workshops (IPDPSW)}, pp.
  \bibinfo{pages}{947--956}.
\bibitem[{Kramer et~al.(2018)Kramer, Noack, Reinefeld, Rodriguez, and
  Zelinskyy}]{Kramer2018a}
\bibinfo{author}{T.~Kramer}, \bibinfo{author}{M.~Noack},
  \bibinfo{author}{A.~Reinefeld}, \bibinfo{author}{M.~Rodriguez},
  \bibinfo{author}{Y.~Zelinskyy},
\newblock \bibinfo{title}{{Efficient calculation of open quantum system
  dynamics and time-resolved spectroscopy with Distributed Memory HEOM
  (DM-HEOM)}},
\newblock \bibinfo{journal}{Journal of Computational Chemistry}
  (\bibinfo{year}{2018}).
\bibitem[{Fenna and Matthews(1975)}]{Fenna1975}
\bibinfo{author}{R.~E. Fenna}, \bibinfo{author}{B.~W. Matthews},
\newblock \bibinfo{title}{{Chlorophyll arrangement in a bacteriochlorophyll
  protein from Chlorobium limicola}},
\newblock \bibinfo{journal}{Nature} \bibinfo{volume}{258}
  (\bibinfo{year}{1975}) \bibinfo{pages}{573--577}.
\bibitem[{Olson(2004)}]{Olson2004}
\bibinfo{author}{J.~M. Olson},
\newblock \bibinfo{title}{{The FMO Protein.}},
\newblock \bibinfo{journal}{Photosynthesis research} \bibinfo{volume}{80}
  (\bibinfo{year}{2004}) \bibinfo{pages}{181--7}.
\bibitem[{Milder et~al.(2010)Milder, Br{\"{u}}ggemann, van Grondelle, and
  Herek}]{Milder2010}
\bibinfo{author}{M.~T.~W. Milder}, \bibinfo{author}{B.~Br{\"{u}}ggemann},
  \bibinfo{author}{R.~van Grondelle}, \bibinfo{author}{J.~L. Herek},
\newblock \bibinfo{title}{{Revisiting the optical properties of the FMO
  protein}},
\newblock \bibinfo{journal}{Photosynthesis Research} \bibinfo{volume}{104}
  (\bibinfo{year}{2010}) \bibinfo{pages}{257--274}.
\bibitem[{May and K{\"{u}}hn(2004)}]{May2004}
\bibinfo{author}{V.~May}, \bibinfo{author}{O.~K{\"{u}}hn},
  \bibinfo{title}{{Charge and Energy Transfer Dynamics in Molecular Systems}},
  \bibinfo{publisher}{Wiley-VCH}, \bibinfo{address}{Weinheim},
  \bibinfo{year}{2004}.
\bibitem[{Mukamel(1995)}]{Mukamel1995}
\bibinfo{author}{S.~Mukamel}, \bibinfo{title}{{Principles of Nonlinear Optical
  Spectroscopy}}, \bibinfo{publisher}{Oxford University Press},
  \bibinfo{address}{Oxford}, \bibinfo{year}{1995}.
\bibitem[{Brixner et~al.(2005)Brixner, Stenger, Vaswani, Cho, Blankenship, and
  Fleming}]{Brixner2005}
\bibinfo{author}{T.~Brixner}, \bibinfo{author}{J.~Stenger},
  \bibinfo{author}{H.~M. Vaswani}, \bibinfo{author}{M.~Cho},
  \bibinfo{author}{R.~E. Blankenship}, \bibinfo{author}{G.~R. Fleming},
\newblock \bibinfo{title}{{Two-dimensional spectroscopy of electronic couplings
  in photosynthesis}},
\newblock \bibinfo{journal}{Nature} \bibinfo{volume}{434}
  (\bibinfo{year}{2005}) \bibinfo{pages}{625--628}.
\bibitem[{Engel et~al.(2007)Engel, Calhoun, Read, Ahn, Man{\v{c}}al, Cheng,
  Blankenship, and Fleming}]{Engel2007a}
\bibinfo{author}{G.~S. Engel}, \bibinfo{author}{T.~R. Calhoun},
  \bibinfo{author}{E.~L. Read}, \bibinfo{author}{T.-K. Ahn},
  \bibinfo{author}{T.~Man{\v{c}}al}, \bibinfo{author}{Y.-C. Cheng},
  \bibinfo{author}{R.~E. Blankenship}, \bibinfo{author}{G.~R. Fleming},
\newblock \bibinfo{title}{{Evidence for wavelike energy transfer through
  quantum coherence in photosynthetic systems}},
\newblock \bibinfo{journal}{Nature} \bibinfo{volume}{446}
  (\bibinfo{year}{2007}) \bibinfo{pages}{782--786}.
\bibitem[{Cho et~al.(2005)Cho, Vaswani, Brixner, Stenger, and
  Fleming}]{Cho2005}
\bibinfo{author}{M.~Cho}, \bibinfo{author}{H.~M. Vaswani},
  \bibinfo{author}{T.~Brixner}, \bibinfo{author}{J.~Stenger},
  \bibinfo{author}{G.~R. Fleming},
\newblock \bibinfo{title}{{Exciton Analysis in 2D Electronic Spectroscopy}},
\newblock \bibinfo{journal}{The Journal of Physical Chemistry B}
  \bibinfo{volume}{109} (\bibinfo{year}{2005}) \bibinfo{pages}{10542--10556}.
\bibitem[{Nuernberger et~al.(2015)Nuernberger, Ruetzel, and
  Brixner}]{Nuernberger2015}
\bibinfo{author}{P.~Nuernberger}, \bibinfo{author}{S.~Ruetzel},
  \bibinfo{author}{T.~Brixner},
\newblock \bibinfo{title}{{Multidimensional Electronic Spectroscopy of
  Photochemical Reactions}},
\newblock \bibinfo{journal}{Angewandte Chemie International Edition}
  \bibinfo{volume}{54} (\bibinfo{year}{2015}) \bibinfo{pages}{11368--11386}.
\bibitem[{Cole et~al.(2013)Cole, Chin, Hine, Haynes, and Payne}]{Cole2013}
\bibinfo{author}{D.~J. Cole}, \bibinfo{author}{A.~W. Chin},
  \bibinfo{author}{N.~D.~M. Hine}, \bibinfo{author}{P.~D. Haynes},
  \bibinfo{author}{M.~C. Payne},
\newblock \bibinfo{title}{{Toward Ab Initio Optical Spectroscopy of the
  Fenna–Matthews–Olson Complex}},
\newblock \bibinfo{journal}{The Journal of Physical Chemistry Letters}
  \bibinfo{volume}{4} (\bibinfo{year}{2013}) \bibinfo{pages}{4206--4212}.
\bibitem[{M{\{}$\backslash$"u{\}}h et~al.(2007)M{\{}$\backslash$"u{\}}h,
  Madjet, Adolphs, Abdurahman, Rabenstein, Ishikita, Knapp, and
  Renger}]{Muh2007}
\bibinfo{author}{F.~M{\{}$\backslash$"u{\}}h}, \bibinfo{author}{M.~E.-A.
  Madjet}, \bibinfo{author}{J.~Adolphs}, \bibinfo{author}{A.~Abdurahman},
  \bibinfo{author}{B.~Rabenstein}, \bibinfo{author}{H.~Ishikita},
  \bibinfo{author}{E.-W. Knapp}, \bibinfo{author}{T.~Renger},
\newblock \bibinfo{title}{{alpha-Helices direct excitation energy flow in the
  Fenna Matthews Olson protein}},
\newblock \bibinfo{journal}{Proceedings of the National Academy of Sciences}
  \bibinfo{volume}{104} (\bibinfo{year}{2007}) \bibinfo{pages}{16862--16867}.
\bibitem[{Adolphs(2008)}]{Adolphs2008a}
\bibinfo{author}{J.~Adolphs}, \bibinfo{title}{{Theory of Excitation Energy
  Transfer in Pigment-Protein Complexes}}, Ph.D. thesis, Freie
  Universit{\"{a}}t Berlin, \bibinfo{year}{2008}.
\bibitem[{Kingma and Ba(2014)}]{Kingma2014}
\bibinfo{author}{D.~P. Kingma}, \bibinfo{author}{J.~Ba},
\newblock \bibinfo{title}{{Adam: A Method for Stochastic Optimization}}
  (\bibinfo{year}{2014}) \bibinfo{pages}{1--15}.
\bibitem[{Hornik et~al.(1989)Hornik, Stinchcombe, and White}]{Hornik1989}
\bibinfo{author}{K.~Hornik}, \bibinfo{author}{M.~Stinchcombe},
  \bibinfo{author}{H.~White},
\newblock \bibinfo{title}{{Multilayer feedforward networks are universal
  approximators}},
\newblock \bibinfo{journal}{Neural Networks} \bibinfo{volume}{2}
  (\bibinfo{year}{1989}) \bibinfo{pages}{359--366}.
\bibitem[{Hochreiter and Schmidhuber(1997)}]{Hochreiter1997}
\bibinfo{author}{S.~Hochreiter}, \bibinfo{author}{J.~Schmidhuber},
\newblock \bibinfo{title}{{Long Short-Term Memory}},
\newblock \bibinfo{journal}{Neural Computation} \bibinfo{volume}{9}
  (\bibinfo{year}{1997}) \bibinfo{pages}{1735--1780}.
\bibitem[{Giles et~al.(2001)Giles, Lawrence, and Tsoi}]{Lawrence1996}
\bibinfo{author}{C.~L. Giles}, \bibinfo{author}{S.~Lawrence},
  \bibinfo{author}{A.~C. Tsoi},
\newblock \bibinfo{title}{{No Title}},
\newblock \bibinfo{journal}{Machine Learning} \bibinfo{volume}{44}
  (\bibinfo{year}{2001}) \bibinfo{pages}{161--183}.
\bibitem[{Lecun et~al.(1998)Lecun, Bottou, Bengio, and Haffner}]{LeCun}
\bibinfo{author}{Y.~Lecun}, \bibinfo{author}{L.~Bottou},
  \bibinfo{author}{Y.~Bengio}, \bibinfo{author}{P.~Haffner},
\newblock \bibinfo{title}{{Gradient-based learning applied to document
  recognition}},
\newblock \bibinfo{journal}{Proceedings of the IEEE} \bibinfo{volume}{86}
  (\bibinfo{year}{1998}) \bibinfo{pages}{2278--2324}.
\bibitem[{Wolfram(2017)}]{Wolfram2018}
\bibinfo{author}{S.~Wolfram}, \bibinfo{title}{{Mathematica 11.3}},
  \bibinfo{year}{2017}.

\end{thebibliography}
\end{document}